\documentclass[runningheads]{llncs}

\usepackage{accv}

\usepackage{accvabbrv}

\usepackage{graphicx}
\usepackage{booktabs}

\usepackage[accsupp]{axessibility}

\usepackage{hyperref}

\usepackage{orcidlink}

\newcommand{\corrauth}{\unskip$^\star$}

\usepackage{float}

\begin{document}

% ---------------------------------------------------------------

\title{SLIM: Semantic-based Low-bitrate Image Compression for Machines}

\author{Hyeonjin Lee\inst{1}\orcidlink{0009-0006-9460-7269} \and
Jun-Hyuk Kim\inst{2}\thanks{Corresponding authors}\orcidlink{0000-0002-7980-4583} \and
Jong-Seok Lee\inst{1}\corrauth\orcidlink{0000-0002-8038-1119}}

\authorrunning{H. Lee et al.}

\institute{School of Integrated Technology / BK21 Graduate Program in Intelligent Semiconductor Technology, Yonsei University, Republic of Korea
\and
Department of Artificial Intelligence, Chung-Ang University, Republic of Korea
\email{\{hyeonjin.lee, jong-seok.lee\}@yonsei.ac.kr} \\
\email{junhyukkim@cau.ac.kr}
}

\maketitle

% ---------------------------------------------------------------
% ---------------------------------------------------------------

\begin{abstract}

In recent years, the demand of image compression models for machine vision has increased dramatically. However, the training frameworks of image compression still focus on the vision of human, preserving unnecessary details and limiting bitrate efficiency for machine vision tasks. In this paper, we propose Semantic-based Low-bitrate Image compression for Machines, termed SLIM. 
This proposed framework consists of two key components: RoI-focused latent compression and semantic latent enhancement with dual cues. 
First, the latent compressor is trained to focus on Region-of-Interest (RoI) areas that are semantically and functionally important for machine vision tasks in the latent space, while requiring no explicit masks during inference.
Second, to compensate for severe information loss at extremely low bitrates, a diffusion-based semantic decoder enhances the compressed latent with a pretrained diffusion model guided by two complementary cues provided by an RoI-focused latent and text-based semantic guidance.
This design preserves RoI details that are critical for machine perception while maintaining semantically coherent and artifact-suppressed non-RoI regions under extremely low bitrates.
Experimental results show that SLIM consistently achieves strong performance across diverse machine vision tasks under the same bits per pixel condition, compared to conventional image compression models for machines.
Our code is available at \url{https://github.com/mcml-official/SLIM}

\end{abstract}

% ---------------------------------------------------------------
% ---------------------------------------------------------------

%%%%%%%%%%%%%%%%%%%%%%%%%%%%%%%%%%%%%%%%%%%%%%%%%%%%%%%%%%%%%%%%%

% ---------------------------------------------------------------
% ---------------------------------------------------------------

\section{Introduction}
\label{sec:intro}

As the demand for image transmission continues to grow, image compression has evolved from traditional codecs such as JPEG  \cite{jpeg} and JPEG2000 \cite{jpeg2000} to learned image compression models \cite{hv_balle_ae, hv_minnen_context} optimized via rate–distortion trade-offs. 
These methods aim to achieve low bitrates while maintaining high quality of reconstructed images.
However, the quality of reconstructed images has been evaluated primarily based on human visual perception, often resulting in suboptimal performance when compressed images are used for machine vision tasks, such as classification, object detection, and semantic segmentation. 
Despite the rapidly increasing proportion of compressed images processed by machines, research addressing this important issue remains limited.

Although several studies \cite{mv_roi, saicm, mpa, transtic, adapticmh} have explored image compression for machine vision, performance at low bitrates remains a major challenge.
This limitation arises because achieving low bitrates through excessive compression requires image compression models to harm pixel-level information across the entire image, thereby degrading essential semantic information and introducing artifacts that impair downstream task performance.

To address this issue, we propose SLIM, \textbf{S}emantic-based \textbf{L}ow-birate \textbf{I}mage compression for \textbf{M}achines by leveraging diffusion.
Our model is designed to better preserve task-relevant information in the Region-of-Interest(RoI) even under low-bitrate conditions, by training an RoI-focused latent compressor.
While several studies \cite{mv_roi, saicm} have investigated RoI-aware compression by assigning different priorities or bit budgets to RoI and non-RoI regions, these approaches typically require an RoI-mask during inference or substantially sacrifice the reconstruction quality of non-RoI regions.
Unlike those approaches, we introduce RoI-focused guidance only into the training loss, using machine-vision-oriented RoI-masks rather than human-vision-oriented masks. This allows the proposed compressor to learn RoI-aware latent representations without requiring any RoI-mask at inference time. Moreover, considering that the quality of non-RoI regions can also affect downstream machine vision task performance, we leverage the generative prior of a pretrained diffusion model to refine compressed latent features, thereby suppressing compression artifacts and reconstructing non-RoI regions with improved texture consistency even under low-bitrate conditions.
To enable the diffusion model to restore images while preserving globally coherent semantics, we additionally inject a text-based semantic guidance.
By jointly exploiting the RoI-focused spatial cues and text-based semantic cues for semantic latent enhancement within a pretrained diffusion pipeline, our method reconstructs high-quality images that preserve task-relevant details and support strong downstream task performance under low-bitrate conditions.

Our main contributions are as follows.
\vspace{-3.3em}

\begin{itemize}

\item We propose a novel image compression framework for machines, SLIM, which learns RoI-focused latent representations to effectively preserve critical semantic information for machine vision under low-bitrate conditions, without requiring explicit RoI-masks during inference time.

\item By introducing a semantic latent enhancement process that refines compressed latents with a dual cues, an RoI-focused spatial cue and a text-based semantic cue, our method guides the employed pretrained diffusion model toward semantically coherent image reconstruction.

\item We demonstrate that our framework generalizes robustly across multiple vision tasks and achieves superior performance compared to other compression methods for machines, especially at low bitrates.

\end{itemize}

% ---------------------------------------------------------------
% ---------------------------------------------------------------

%%%%%%%%%%%%%%%%%%%%%%%%%%%%%%%%%%%%%%%%%%%%%%%%%%%%%%%%%%%%%%%%%

% ---------------------------------------------------------------
% ---------------------------------------------------------------

\section{Related Works}
\label{sec:related}

\vspace{-0.4em}
\subsection{Learned Image Compression}
Traditional image compression methods \cite{jpeg, jpeg2000, bpg, vvc}, including JPEG, JPEG2000, BPG, and VVC intra coding, rely on carefully designed coding tools such as transform or predictive coding, quantization, and entropy coding. 
These hand-engineered pipelines have been extensively studied and optimized in conventional compression systems.
Although conventional codecs rely on well-optimized encoders, their overall coding pipelines are manually designed rather than learned end-to-end from image data.

In contrast, learned image compression (LIC) methods \cite{hv_balle_ae} use neural networks to model key components of the compression pipeline and train the resulting models end-to-end with a rate-distortion objective.
Thus these methods have demonstrated strong rate-distortion performance compared with conventional codec baselines. 
Their typical training objective is defined as a rate-distortion trade-off:
{\setlength{\abovedisplayskip}{3pt}
 \setlength{\belowdisplayskip}{3pt}
 \setlength{\abovedisplayshortskip}{0pt}
 \setlength{\belowdisplayshortskip}{3pt}
\begin{equation}
    \mathcal{L} = \mathcal{R} + \lambda \cdot \mathcal{D}(x, \hat{x}),
\label{eqn:loss_lic}
\end{equation}
}
\noindent where $x$ and $\hat{x}$ denote the original and reconstructed images, respectively, $\mathcal{D}(x, \hat{x})$ is a distortion loss measuring reconstruction quality, $\mathcal{R}$ is the estimated bitrate, and $\lambda$ balances the trade-off between rate and distortion.
Building upon this formulation, recent works have enhanced entropy modeling and representation capacity through hyperprior models, autoregressive context models, and transformer-based architectures \cite{hv_balle2_hyper, hv_minnen_context, hv_transformer-based1}. 

To further improve reconstruction quality under limited bit budgets, recent LIC studies have explored several extensions, including RoI-aware bit allocation and generative reconstruction:

\noindent\textbf{RoI-aware image compression.}
Beyond uniform image compression, several studies \cite{mv_roi, saicm} have explored RoI-aware compression to preserve semantically important or task-relevant regions more effectively under limited bit budgets. 
These methods typically allocate more bits to RoI areas or impose stronger reconstruction constraints on them. 
While such strategies can improve the quality of important regions, they often depend on externally provided RoI-masks even during inference, or sacrifice the reconstruction quality of non-RoI regions when the bitrate becomes very low.

\noindent\textbf{Generative image compression.}
Another line of work \cite{hv_ganbase, resulic, perco, lic_diff_conditional, oscar} incorporates generative models, such as GAN \cite{gan} and diffusion models \cite{diffusion, ldm}, into image compression. 
By synthesizing perceptually plausible details, these methods can improve visual realism, especially at low bitrates where conventional reconstructions tend to be blurry or over-smoothed. 
However, maintaining faithful reconstruction remains challenging under extremely low-bitrate conditions, as generated details may be visually plausible but not semantically faithful to the original image.

Although RoI-aware and generative compression improve low-bitrate reconstruction in different ways, severe quality degradation can still occur when the bit budget becomes extremely limited. 
RoI-aware compression can emphasize important regions, but excessive degradation of surrounding regions may remove or distort contextual cues that are useful for visual analysis. 
Generative compression can recover visually plausible details, but under extremely low bitrates, insufficient transmitted information may lead to distorted structures or unreliable local details. 
Such degradation is undesirable for human viewing and can be even more critical for machine vision, where downstream models are sensitive to local distortions, missing context, and structural inconsistencies. 
Motivated by this issue, we investigate diffusion-based image compression for machine vision, focusing on machine-oriented reconstruction that preserves semantically important regions, contextual structures, and overall image coherence under extremely limited bit budgets.

\subsection{Image Compression for Machine Vision}
The image compression models trained for human vision have the issue that their reconstructed images are suboptimal for machine vision, thus showing a significant drop in performance when the reconstructed images are used in the downstream tasks including classification, object detection, and instance segmentation.
To handle this issue, some studies have focused on the image compression models for machine vision.

Some approaches \cite{mv_roi, saicm} sacrifice the image quality, discarding background regions or features which are not necessary for machine vision, in order to improve compression efficiency.
While effective for bitrate reduction, such selective preservation may unintentionally degrade semantic cues outside the selected regions, although these cues can still contribute to downstream recognition.
Another approach \cite{mpa}, which trains a model for both human vision and machine vision simultaneously, achieves high fidelity while reconstructing the image suitable for machine vision but requires additional cost for training multiple encoders.
There are other approaches \cite{transtic, adapticmh} that convert pretrained image compression models for human vision into those for machine vision.
Beyond image-based methods, related objectives have also been explored in video semantic compression \cite{video_compression1, video_compression2}.
Differently from these existing studies, our study utilizes a pretrained latent diffusion model to construct an improved image compression model for machine vision.
And our framework aggressively removes unnecessary perceptual details required only for human vision, but still reconstructs high quality images by using the diffusion model.

% ---------------------------------------------------------------
% ---------------------------------------------------------------

%%%%%%%%%%%%%%%%%%%%%%%%%%%%%%%%%%%%%%%%%%%%%%%%%%%%%%%%%%%%%%%%%

% ---------------------------------------------------------------
% ---------------------------------------------------------------

\section{Proposed Method}
\label{sec:proposed}

\begin{figure}[t]
  \centering
  \vspace{-1em}
  \includegraphics[width=1.0\linewidth]{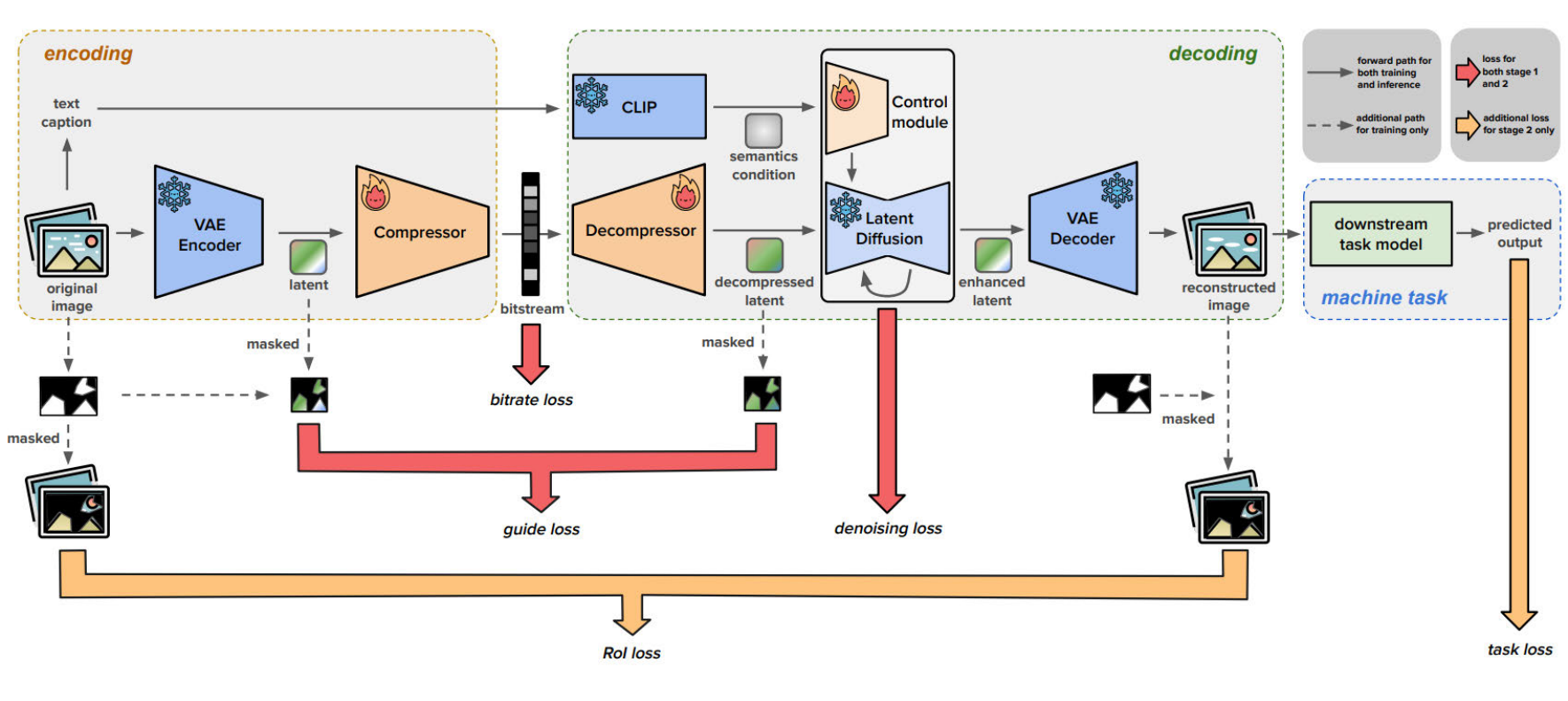}
  \vspace{-3em}

  \caption{Architecture of the proposed SLIM. Our model employs a two-stage training strategy; the red arrows mean the losses used in both stages, and the orange arrows indicate the additional losses used in stage 2.}

   \label{fig: 2_flow_of_training}
      \vspace{-1.5em}

\end{figure}

\vspace{-0.4em}
\subsection{System Architecture}
The overall architecture of our model is shown in Fig. \ref{fig: 2_flow_of_training}. 
The encoder part consists of a VAE encoder and a compressor, while the decoder part includes a decompressor, a latent diffusion model with a control module, a CLIP encoder, and a VAE decoder. We leverage pretrained components from Stable Diffusion 2.1 \cite{ldm}, and train only the compression-related and control components.
Additional details on the compressor and decompressor structures are provided in the supplementary material.

At the encoder side, the input image is first mapped to a latent representation by the VAE encoder, as operating in the latent space enables efficient learned compression.
The latent is then compressed into a bitstream by a hyperprior-based compressor inspired by ELIC \cite{elic}, which employs a latent analysis transform and a hyper analysis transform to generate a compressed bitstream. 
During training, an RoI-mask is used to guide the compressor to implicitly allocate more bits to semantically important regions (\textbf{RoI-focused latent compression}).
Importantly, no RoI information is provided during inference, and the trained compressor operates solely on the latent input.

At the decoder side, the received bitstream is first decoded by the decompressor, which follows the corresponding hyperprior-based synthesis architecture of ELIC \cite{elic} and reconstructs a decompressed latent via a hyper synthesis transform and a latent synthesis transform.
Instead of directly feeding the decompressed latent into the VAE decoder, we first enhance it into a semantically enriched latent representation that is optimized to generate high-quality images for downstream machine tasks (\textbf{Semantic latent enhancement with dual cues}).
Specifically, the decompressed latent representation is refined by a UNet-based latent diffusion model conditioned through a control module, using two complementary cues derived from the RoI-focused latent and from text-based semantic guidance.
The control module follows the ControlNet design \cite{controlnet}, adopting the same UNet backbone architecture as the latent diffusion model and modulating its intermediate features through zero-initialized convolution layers.
A text caption generated from the input image is encoded by a frozen CLIP encoder and injected into the diffusion model to guide the denoising process.
Finally, the enhanced latent is decoded by the VAE decoder to produce the reconstructed image, which is directly used for downstream machine vision tasks.

%%%%%%%%%%%%%%%%%%%%%%%%%%%%%%%%%%%%%%%%%%%%%%%%%%%%%%%%%%%%%%%%%%%%%%%%%%%%%%%%

\subsection{RoI-focused Latent Compression}
\label{subsec: roi-focused latent compression}

To effectively transmit task-critical information to machine vision systems under extremely constrained bit budgets, our framework adopts an RoI-focused latent compression strategy. 
Unlike many RoI-based image compression methods that rely on human-centric object masks, such as instance masks or segmentation maps, our RoI-mask is defined from a machine-vision perspective. 
Specifically, for each training image, we compute a Grad-CAM \cite{gradcam} heatmap from a machine task model, using a pretrained ResNet-50 \cite{resnet} in this work.
The heatmap is then normalized to $[0, 1]$ and binarized using a threshold of $0.5$ to generate the RoI-mask.
The resulting RoI-mask identifies regions that are important for downstream machine perception, without necessarily covering entire object regions. 
Note that the RoI-mask can also include task-relevant contextual regions, such as semantically informative background regions, to support scene understanding. 
Thus, it serves as a machine-oriented importance map that guides the latent compressor to allocate more representational capacity to task-critical visual information. 

The RoI-mask is used only to define the training objectives and is not provided as an input to either the compressor or the decoder. 
During training, we minimize the discrepancy between the uncompressed latent representation and its decompressed counterpart within the RoI, encouraging the decompressed latent representation to preserve RoI-relevant information. 
This objective is computed only over the RoI-mask that is resized to the latent resolution. 
We further impose an auxiliary RoI reconstruction objective to enforce pixel-level and semantic-level fidelity between the original and reconstructed images, with the corresponding losses computed only within the RoI-mask at the image resolution.
The detailed formulations of the training objectives are presented in Section \ref{subsec: training phase}. 
With these mask-guided objectives, the RoI-focused latent compression behavior is encoded into the trained SLIM model parameters. 
Thus, the compressor performs mask-free inference using only the input image. 
In Section \ref{sec:experiment_ablation}, the bit allocation map presented in the ablation study on the RoI-focused latent empirically shows that the compressor trained within our proposed framework learns to allocate more bits to RoI regions, without requiring the RoI-mask at inference time.

By prioritizing machine-critical RoI regions rather than uniformly preserving the entire image under a fixed bit budget, our method retains informative visual cues even at extremely low bitrates. 
However, although non-RoI regions are not the primary target of preservation during compression, severely degraded non-RoI content can introduce unnatural artifacts and distortions that disrupt the semantic consistency of the reconstructed image and may adversely affect downstream machine perception. 
Therefore, we introduce a semantic latent enhancement process, which is detailed in the following section.

\subsection{Semantic Latent Enhancement with Dual Cues}
\label{subsec: semantic-based diffusion decoder}

Under extremely low-bitrate compression, the transmitted latent often lacks sufficient semantic and structural information for reliable reconstruction. 
To mitigate this information loss, we introduce a semantic latent enhancement process that refines the decompressed latent using a pretrained latent diffusion model guided by two complementary cues: an RoI-focused spatial cue and a text-based semantic cue.
The RoI-focused spatial cue is derived from the RoI-focused latent. Since this latent is obtained through a compressor–decompressor pair and preserves fine-grained structural details within RoI regions, the resulting cue provides RoI-localized spatial guidance.
The text-based semantic cue, on the other hand, is derived from the text-based semantic guidance. Since this guidance is constructed by encoding a textual description of the input image with CLIP \cite{clip}, the resulting cue serves as global scene-level semantic guidance. In our experiments, we obtain the textual description using BLIP2 \cite{blip2}.
These two cues complement each other by preserving local RoI structures while enforcing high-level semantic consistency during the diffusion process.

For semantic latent enhancement, our diffusion-based decoder consists of a frozen latent diffusion backbone and a trainable control module. 
We condition the denoising process on dual cues,
{\setlength{\abovedisplayskip}{3pt}
 \setlength{\belowdisplayskip}{3pt}
 \setlength{\abovedisplayshortskip}{0pt}
 \setlength{\belowdisplayshortskip}{3pt}
\begin{equation}
    \mathcal{S}_{\text{dual}} = (\hat{z}, c),
    \label{eq:dual_semantic_cues}
\end{equation}
}
where $\hat{z}$ denotes the RoI-focused latent reconstructed by the decompressor, and $c$ denotes the CLIP-encoded text-based semantic guidance.
Specifically, $\hat{z}$ is injected into the control module as a latent hint, allowing the decoder to retain RoI-aware spatial structures. 
Meanwhile, $c$ is provided as the cross-attention context, regularizing the denoising process toward semantically consistent content. 
Accordingly, given the noisy latent $z_t$, timestep $t$, and dual cues $\mathcal{S}_{\text{dual}}$, the decoder predicts the added noise $\epsilon$ under both spatial and semantic guidance. 
This dual conditioning enables the diffusion-based decoder to recover an enhanced latent representation that preserves RoI-specific details while suppressing unnatural artifacts across the entire scene, thereby reducing interference with downstream machine vision tasks.
The enhanced latent representation is finally decoded into the image domain by the VAE decoder.

%%%%%%%%%%%%%%%%%%%%%%%%%%%%%%%%%%%%%%%%%%%%%%%%%%%%%%%%%%%%%%%%%%%%%%%%%%%%%%%%%%%%%%%%%%

\subsection{Two-Stage Training}
\label{subsec: training phase}
As shown in Fig. \ref{fig: 2_flow_of_training}, the training phase of SLIM is divided into two stages.
The training stage 1 emphasizes compact representation by removing redundant information, while the training stage 2 additionally focuses on preserving and restoring essential details to improve reconstruction quality and task performance, which is non-trivial to achieve in a single stage.
Details are as follows.

\subsubsection{Stage 1} The trainable components of SLIM are optimized using a combination of bitrate loss, guide loss, and denoising loss.
The bitrate loss encourages efficient compression, the guide loss enforces high-fidelity reconstruction in RoI regions, and the denoising loss trains the control module to guide the frozen diffusion backbone in accurately predicting noise conditioned on the dual cues.

We define the guide loss as follows:
\begin{equation}
    \mathcal{L}_{\text{guide}}
    =
    \frac{1}{\left\|
    \mathcal{M}_{\text{ROI}}
    \right\|_1}
    \left\|
    \mathcal{M}_{\text{ROI}} \odot (z - \hat{z})
    \right\|_2^2 
    \label{eqn:loss_guide}
\end{equation}
\noindent where $z$ and $\hat{z}$ mean the latent encoded by the VAE encoder and the latent decompressed by the decompressor, respectively.
$\mathcal{M}_{\text{ROI}}$ denotes the binary RoI-mask, which is resized and broadcast to match the shape of the corresponding input.
Thus, $\|\mathcal{M}_{\text{ROI}}\|_1$ denotes the number of valid elements after applying the RoI-mask in each domain, and $\odot$ denotes element-wise multiplication.
Combined with the bitrate loss $\mathcal{R}$, SLIM is trained to reduce more bits on non-RoI areas where fidelity is less critical.

Besides, we define the denoising loss as follows:
\begin{equation}
    \mathcal{L}_{\text{denoise}}
    =
    \mathbb{E}_{z, \hat{z}, c, t, \epsilon}
    \left[
    \frac{1}{N}
    \left\|
    \epsilon - \epsilon_{\theta}(z_t, t, \mathcal{S}_{\text{dual}})
    \right\|_2^2
    \right]
    \label{eqn:loss_denoise}
\end{equation}

\noindent where $\epsilon$ is the target noise and $\epsilon_{\theta}$ is the predicted noise by the latent diffusion model attached with the control module with weight parameters $\mathcal{\theta}$. 
$N$ denotes the number of elements in each latent representation.
Conditioned on the dual cues $\mathcal{S}_{\text{dual}}$, the diffusion-based decoder learns to perform semantic latent enhancement.

Thus, the loss function of the training stage 1 is as follows:
{\setlength{\abovedisplayskip}{3pt}
 \setlength{\belowdisplayskip}{3pt}
 \setlength{\abovedisplayshortskip}{0pt}
 \setlength{\belowdisplayshortskip}{3pt}
\begin{equation}
    \mathcal{L}_{stage 1} = \lambda_{r} \cdot \mathcal{R} + \lambda_{g} \cdot \mathcal{L}_{guide} +\lambda_{d} \cdot \mathcal{L}_{denoise}
    \label{eqn:loss_stage1}
\end{equation} 
}
\noindent where $\lambda_{r}$, $\lambda_{g}$, and $\lambda_{d}$ are hyperparameters to balance the losses.

\subsubsection{Stage 2} 
Based on $\mathcal{L}_{stage 1}$, additional losses are included in the training stage 2.
To ensure the reconstructed image samples maintain the fidelity and semantics of RoI areas in original images, we define the RoI loss as follows: 
\begin{equation}
\begin{split}
\mathcal{L}_{\text{ROI}}
&=
% \frac{1}{N_{\text{ROI}}}
\frac{1}{\left\|
\mathcal{M}_{\text{ROI}}
\right\|_1}
\left\|
\mathcal{M}_{\text{ROI}} \odot (x - \hat{x})
\right\|_2^2 \\
&\quad
+ \lambda_l \,
\mathrm{LPIPS}
\left(
\mathcal{M}_{\text{ROI}} \odot x,
\mathcal{M}_{\text{ROI}} \odot \hat{x}
\right),
\end{split}
\label{eq:loss_roi}
\end{equation}
\noindent where $\mathrm{LPIPS}(\cdot,\cdot)$ means the LPIPS loss \cite{lpips} between two inputs, and $\lambda_{l}$ means a hyperparameter to balance the two terms of the RoI loss.
The first term enforces pixel-level fidelity between the original image $x$ and reconstructed image $\hat{x}$ in the RoI areas, while the second term enforces high perceptual details of the RoI area in the reconstructed image at the semantic level.

To ensure the reconstructed image samples are optimized for machine vision, we use an additional task loss $\mathcal{L}_{task}$.
Among various downstream tasks, we adopt image classification as a representative task during training and define $\mathcal{L}_{task}$ as the cross-entropy loss between the ground-truth label of the original image and the predicted label obtained from the reconstructed image. 
Since image classification captures high-level semantic concepts, optimizing this loss encourages the reconstructed images to preserve semantic information that is broadly useful across diverse vision tasks, including object detection and instance segmentation.

Thus, the loss function of the training stage 2 is written as:
{\setlength{\abovedisplayskip}{3pt}
 \setlength{\belowdisplayskip}{3pt}
 \setlength{\abovedisplayshortskip}{0pt}
 \setlength{\belowdisplayshortskip}{3pt}
\begin{equation}
    \mathcal{L}_{stage 2} = \mathcal{L}_{stage 1} + \lambda_{ROI} \cdot \mathcal{L}_{ROI} +\lambda_{t} \cdot \mathcal{L}_{task}
    \label{eqn:loss_stage2}
\end{equation} 
}
\noindent where $\lambda_{ROI}$ and $\lambda_{t}$ are hyperparameters to balance the losses.

% ---------------------------------------------------------------
% ---------------------------------------------------------------

%%%%%%%%%%%%%%%%%%%%%%%%%%%%%%%%%%%%%%%%%%%%%%%%%%%%%%%%%%%%%%%%%

% ---------------------------------------------------------------
% ---------------------------------------------------------------

\section{Experimental Results}
\label{sec:experiment}

\vspace{-0.4em}
\subsection{Experimental Setup}

We evaluate SLIM against pretrained image compression models from prior work.
And further, we analyze how each component of SLIM affects the performance of the model through ablation studies.
We also measure computational costs.
Details of training settings, datasets, and compared models are provided below.

\noindent \textbf{Training setup.}
We train our SLIM model for 100,000 steps in each of the two stages, using the AdamW optimizer with the learning rate 0.0001.
We randomly select 3,000 images from the ImageNet-train \cite{imagenet} dataset, then construct a training dataset consisting of images, RoI-masks, and text captions.
The caption bitrate is negligible ($\approx 0.0014$ bpp after zlib lossless compression) and thus excluded from bpp calculations.
Example training samples are provided in the supplementary material.

\noindent \textbf{Loss weight setting.}
In all experiments, the loss weights introduced in Eqs. (\ref{eqn:loss_stage1})-(\ref{eqn:loss_stage2}) are fixed across settings, except for the rate weight $\lambda_{r}$, which is adjusted according to the target bitrate.
Losses related to RoI-focused latent compression and semantic latent enhancement are treated as primary objectives and thus assigned with unit weights ($\lambda_{g} = 1$, $\lambda_{d} = 1$, $\lambda_{l} = 1$, $\lambda_{ROI} = 1$).
We vary the bitrate loss weight $\lambda_{r}$ from $0.03$ to $24.0$ in Stage 1 and from $0.035$ to $28.0$ in Stage 2, using a larger range in Stage 2 to maintain the contribution of the rate term when additional loss terms are introduced.
The task loss weight is set to $\lambda_{t} = 0.05$ to incorporate task-related supervision while preserving stable optimization of the compression objectives.

\noindent \textbf{Environment.}
All experiments are conducted on a single NVIDIA GeForce RTX 4090 GPU, using Python 3.10, PyTorch 2.0.1 with CUDA 11.7.

\noindent \textbf{Comparison.} 
The pretrained models from prior studies used as comparison models are as follows: SA-ICM \cite{saicm}, trained from scratch for machine vision; TransTIC \cite{transtic} and AdaptICMH \cite{adapticmh}, transferred or adapted for machine vision from the base models trained for human vision; and MPA \cite{mpa}, trained to be applicable to both human vision and machine vision tasks.
We also use the pretrained model of OSCAR \cite{oscar}, a state-of-the-art diffusion-based model trained for human vision.
We evaluate the rate–accuracy performance on the ImageNet-val dataset for classification and the rate–mAP performance on the COCO2017-val \cite{coco} dataset for object detection and instance segmentation. Classification performance is measured using ResNet50 \cite{resnet} and ConvNeXt-T \cite{convnext}, while Faster R-CNN \cite{fasterrcnn} and Mask R-CNN \cite{maskrcnn} are used for object detection and instance segmentation, respectively.

\begin{figure}[t]
    \centering

    \begin{subfigure}{0.48\linewidth}
        \centering
        \includegraphics[width=\linewidth]{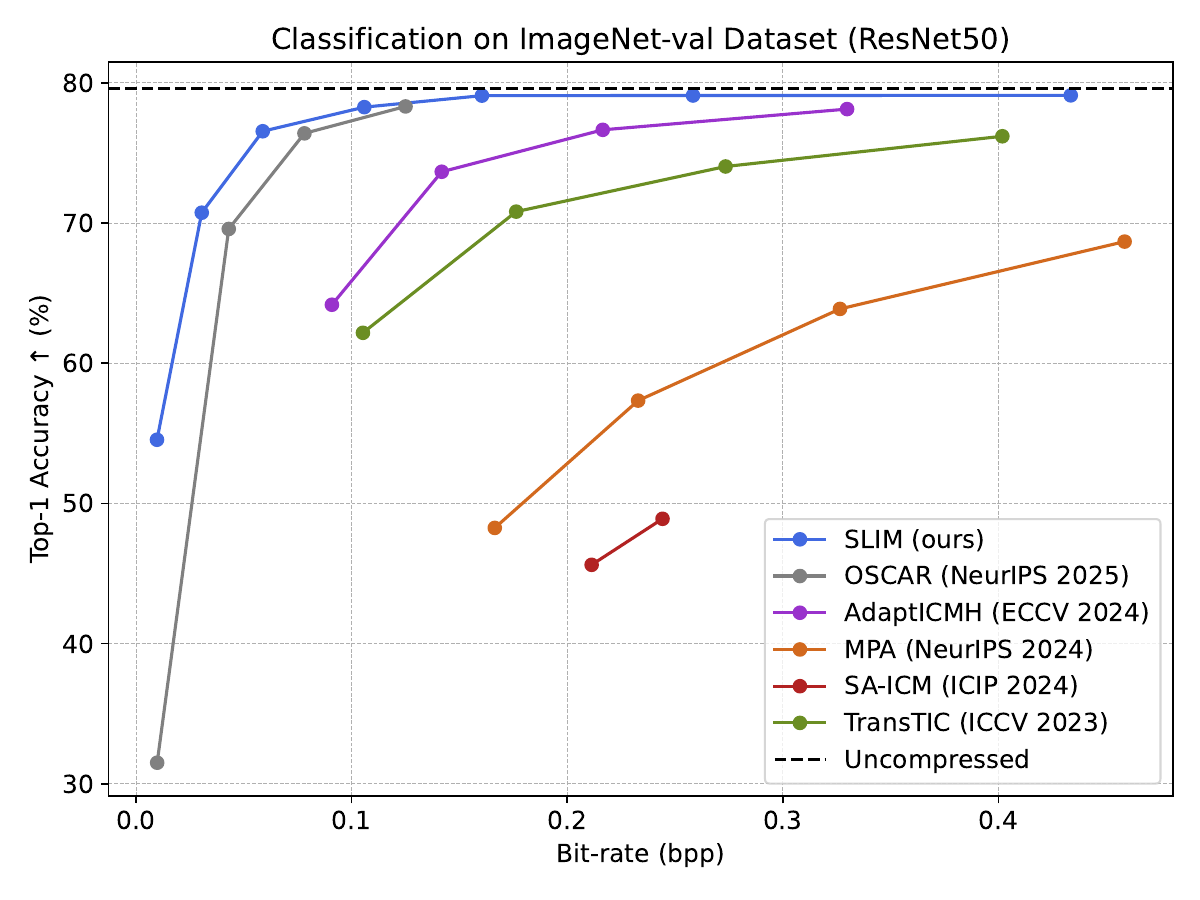}
        \vspace{-1.5em}
        \caption{\centering BPP vs. Accuracy \\(Classification, ResNet50)}
        \label{fig:resnet}
    \end{subfigure}
    \hfill
    \begin{subfigure}{0.48\linewidth}
        \centering
        \includegraphics[width=\linewidth]{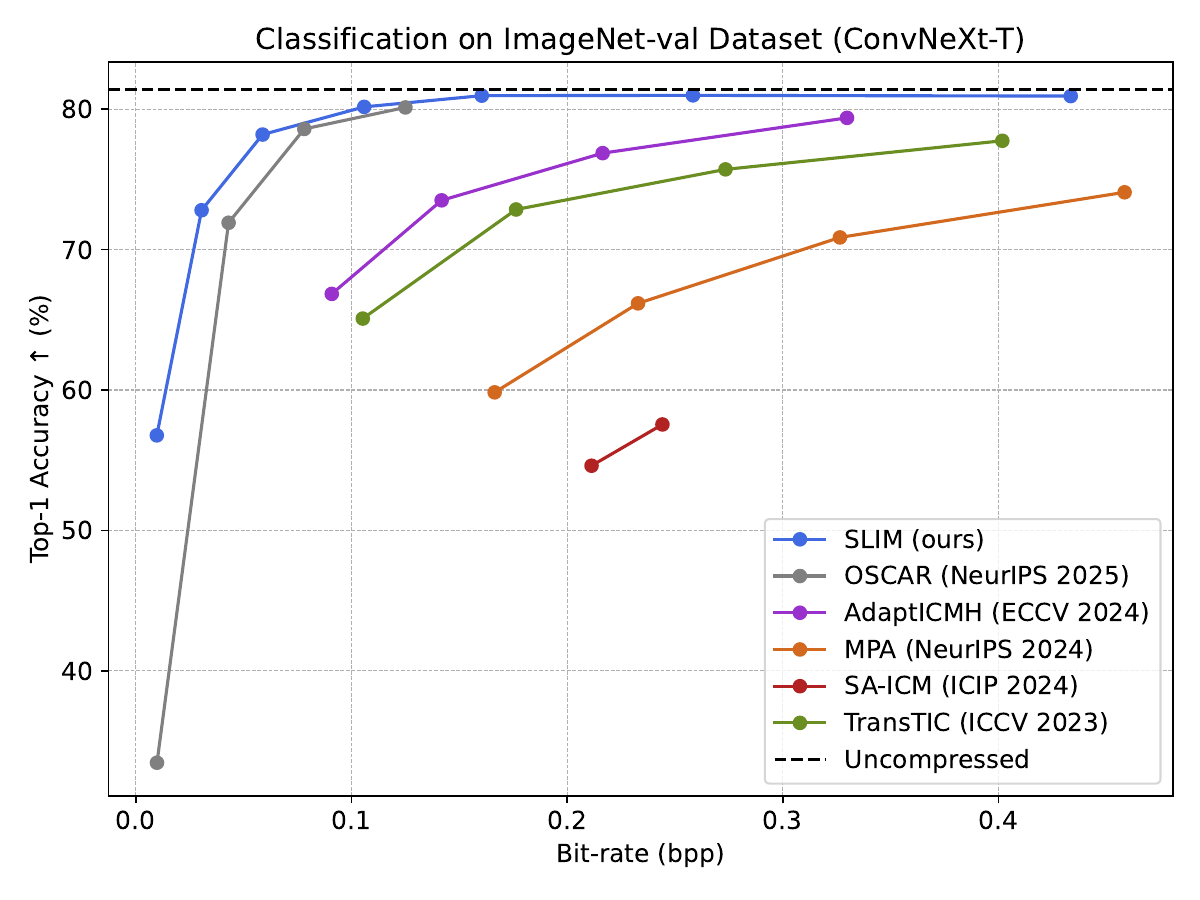}
        \vspace{-1.5em}
        \caption{\centering BPP vs. Accuracy \\(Classification, ConvNeXt-T)}
        \label{fig:convnext}
    \end{subfigure}

    \vspace{0.5em}

    \begin{subfigure}{0.48\linewidth}
        \centering
        \includegraphics[width=\linewidth]{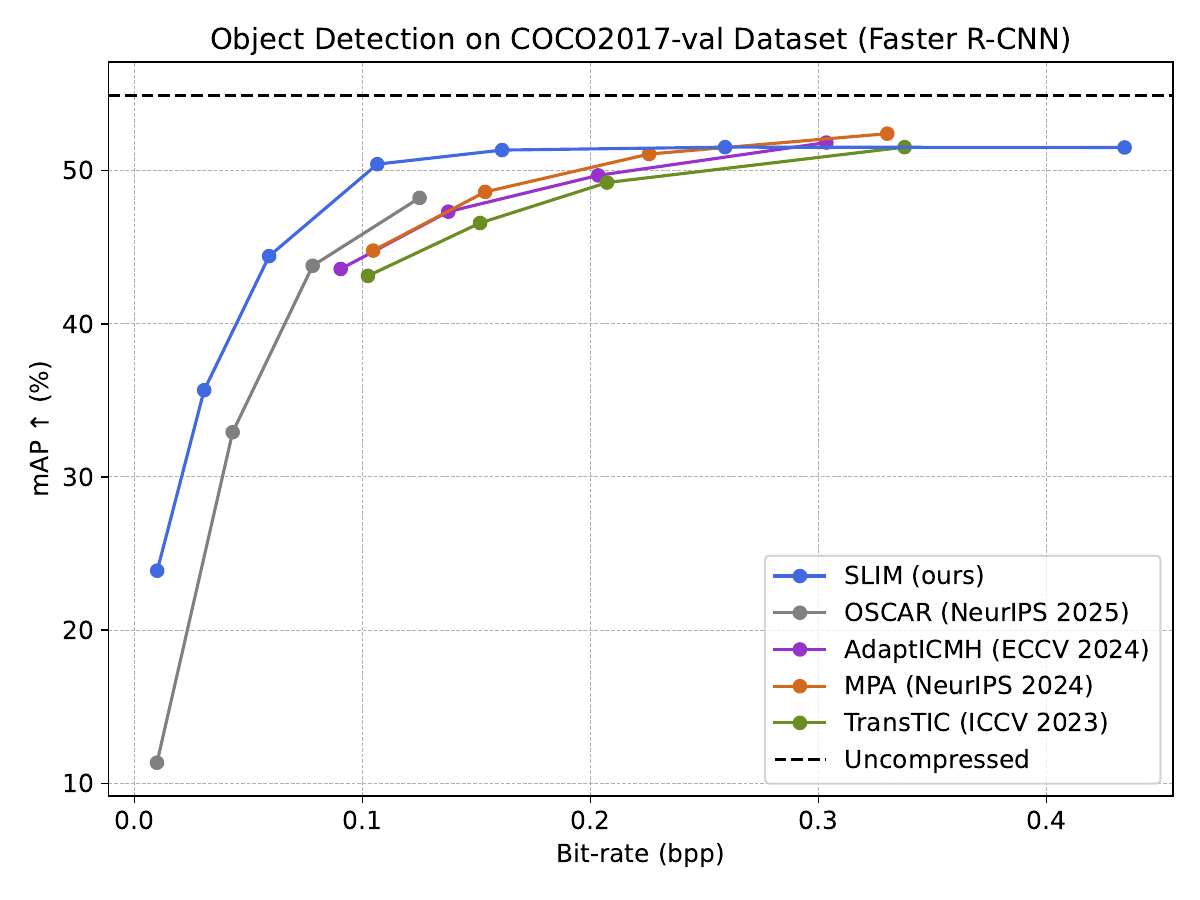}
        \vspace{-1.5em}
        \caption{\centering BPP vs. mAP \\(Object detection, Faster R-CNN)}
        \label{fig:object}
    \end{subfigure}
    \hfill
    \begin{subfigure}{0.48\linewidth}
        \centering
        \includegraphics[width=\linewidth]{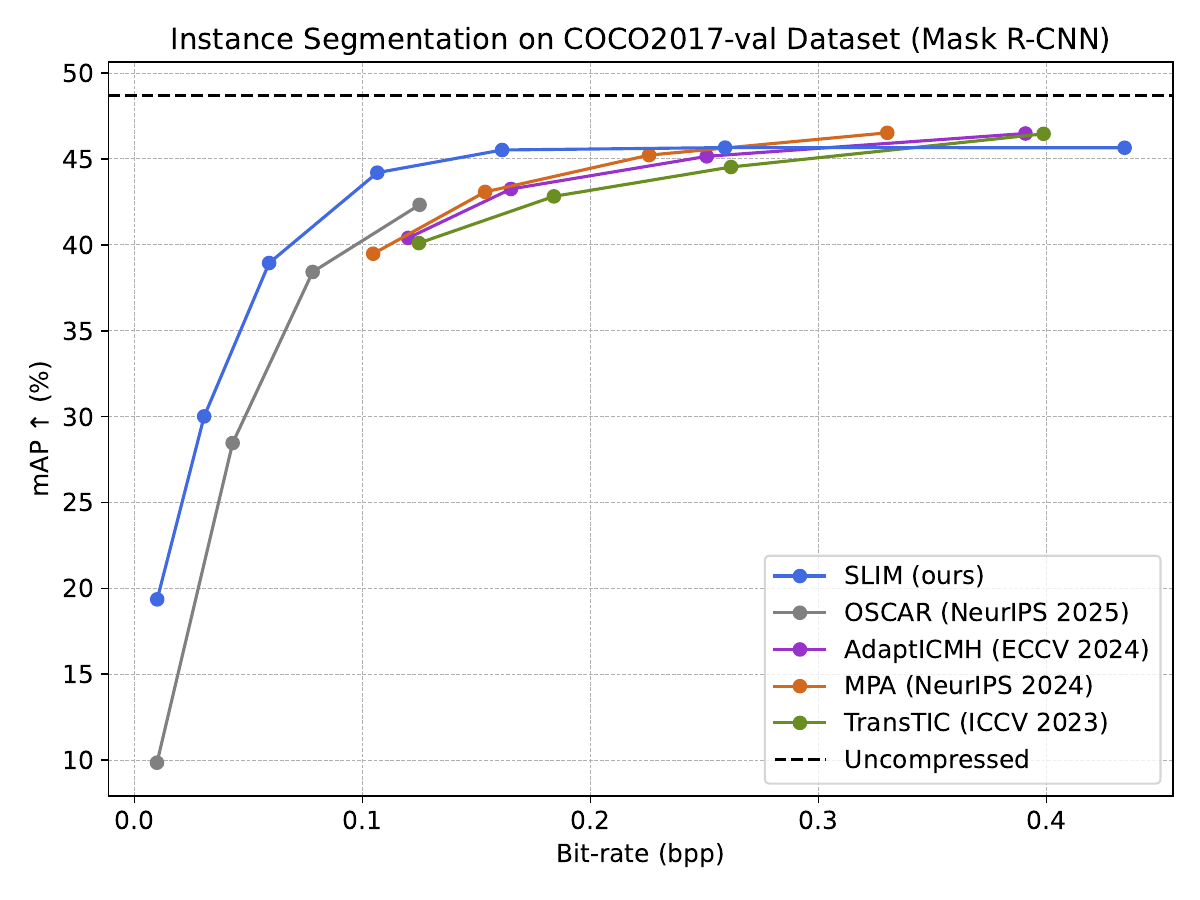}
        \vspace{-1.5em}
        \caption{\centering BPP vs. mAP \\(Instance seg., Mask R-CNN)}
        \label{fig:segment}
    \end{subfigure}

    \vspace{-0.5em}
    \caption{
    Rate-task performance comparison. Classification accuracy is evaluated on the ImageNet-val dataset using (a) ResNet50 and (b) ConvNeXt-T. Mean average precision (mAP) at an IoU threshold of 0.5 is evaluated on the COCO2017-val dataset using (c) Faster R-CNN for object detection and (d) Mask R-CNN for instance segmentation.
    }
    \label{fig: 5_bpp_vs_acc}

    \vspace{-1em}
\end{figure}

\subsection{Rate-Task Performance}

As shown in Fig. \ref{fig: 5_bpp_vs_acc} (a) and (b), our SLIM model achieves the best rate-accuracy performance compared to the other baselines for both ResNet50 and ConvNeXt-T classifiers.
Considering that SLIM is trained using RoI-masks extracted from the pretrained ResNet50 in this study, the strong performance observed with the ConvNeXt-T classifier indicates that our method generalizes well on different classifier architectures.
In comparison to the models for machine vision, including MPA, TransTIC, and AdaptICMH, our approach requires consistently fewer bits per pixel (bpp) to achieve the same classification accuracy. 
Furthermore, in comparison to the model for human vision, OSCAR, our approach still achieves better rate-accuracy performance, particularly at low bitrates.
This suggests that although both SLIM and OSCAR utilize the diffusion model, SLIM is specially designed for machine vision and its effectiveness is increasingly noticeable when the compression ratio increases.

Beyond image classification, we further evaluate the proposed method on downstream vision tasks other than classification to assess task-level generalization.
As shown in Fig. \ref{fig: 5_bpp_vs_acc} (c) and (d), our SLIM model consistently achieves competitive rate-mAP performance across object detection and instance segmentation when evaluated with Faster R-CNN and Mask R-CNN, respectively.
In all cases, the performance improvement by our method is particularly prominent in low-bitrate regimes.

The BD-Rate comparison results are summarized in Table \ref{tab: 1_bpp_vs_acc}.
We compute BD-Rate following Bjøntegaard’s method \cite{bdrate}, where a negative value indicates higher coding efficiency compared to the reference model, for which we use MPA \cite{mpa}. 
SLIM consistently achieves substantial BD-Rate reductions with both ResNet50 and ConvNeXt-T classifiers. 
In Table \ref{tab: 1_extra_bpp_vs_coco}, similar gains are observed for object detection and instance segmentation, demonstrating strong generalization across vision tasks. 
These results indicate that our RoI-focused latent compression effectively preserves semantically important regions for machine vision regardless of the task model.

\begin{table}[t]
  \caption{BD-Rate comparison on the ImageNet-val dataset using ResNet50 and ConvNeXt-T for image classification.}
  \vspace{-0.5em}
  \label{tab: 1_bpp_vs_acc}
  \centering
  \scalebox{1.0}{
  \begin{tabular}{lcc}
    \hline
    \textbf{Model} & \textbf{ResNet50 (\%) $\downarrow$} & \textbf{ConvNeXt-T (\%) $\downarrow$}\\
    \hline
    MPA \cite{mpa} & - & -\\
    TransTIC \cite{transtic} & -67.23 & -55.16 \\
    AdaptICMH \cite{adapticmh} & -75.60 & -65.14 \\
    OSCAR \cite{oscar} & -90.89 & -87.26 \\
    \textbf{SLIM (Ours)} & \textbf{-94.95} & \textbf{-92.68} \\
    \hline
  \end{tabular}
  }
\end{table}

\begin{table}[t]
  \caption{BD-Rate comparison on the COCO2017-val dataset using
Faster R-CNN for object detection and
Mask R-CNN for instance segmentation.}
    \vspace{-0.5em}

  \label{tab: 1_extra_bpp_vs_coco}
  \centering
  \scalebox{1.0}{
  \begin{tabular}{lcc}
    \hline
    \textbf{Model} & \textbf{Faster R-CNN (\%) $\downarrow$} & \textbf{Mask R-CNN (\%) $\downarrow$}\\
    \hline
    MPA \cite{mpa} & - & - \\ %%%% for MPA base
    TransTIC \cite{transtic} & 25.40 & 25.22 \\

    AdaptICMH \cite{adapticmh} & 7.75 & 8.17 \\

    OSCAR \cite{oscar} & -16.15 & -12.94 \\

    \textbf{SLIM (Ours)} & \textbf{-43.65} & \textbf{-40.53} \\
    \hline
    
  \end{tabular}
  }

  \vspace{-1em}

\end{table}

\begin{figure}[t]
  \centering
  \includegraphics[width=1.0\linewidth]{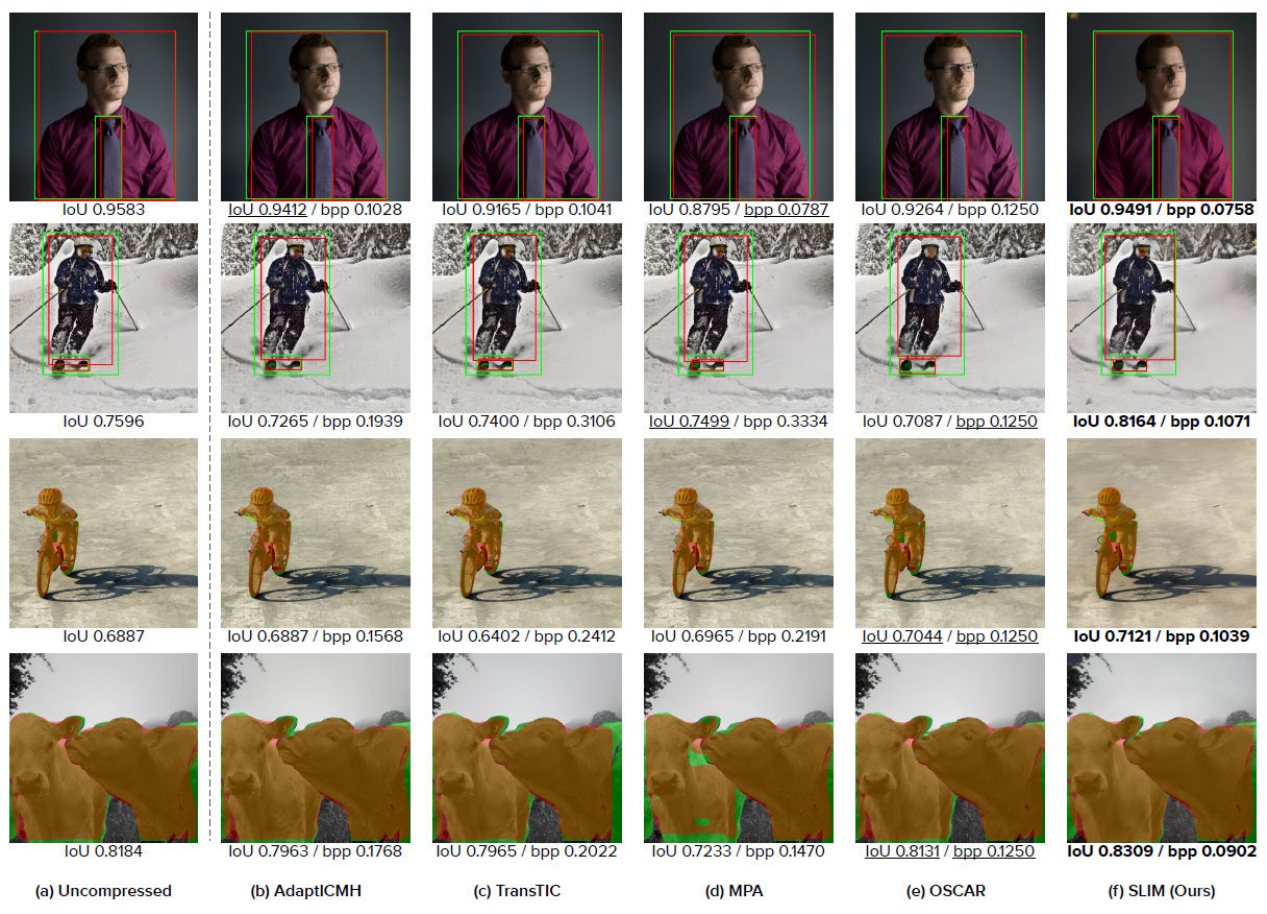}
  \vspace{-2em}

   \caption{Reconstructed samples from different models: AdaptICMH \cite{adapticmh}, TransTIC \cite{transtic}, MPA \cite{mpa}, OSCAR \cite{oscar}, and Ours. Green bounding boxes and masks denote the ground-truth (GT) annotations, while red bounding boxes and masks indicate the model predictions for object detection and instance segmentation, respectively. The IoU and BPP values are shown below each image. The best results are in bold, and the second-best results are underlined.}
   \label{fig: 6_samples}
   \vspace{-1.5em}

\end{figure}

\subsection{Qualitative Results}

We further provide qualitative comparisons with other LIC models for machine vision (AdaptICMH, TransTIC, MPA) and for human vision (OSCAR). 
As shown in Fig. \ref{fig: 6_samples}, our method preserves more faithful and natural textures in RoI regions for machine vision at comparable bpp levels. 
In contrast, other methods often blur or discard fine details.
These results are consistent with the improved rate-task performance of SLIMs.
Rather than uniformly preserving details, our method keeps RoI-focused information in the latent representation, while the diffusion-based decoder enhances the overall reconstruction to avoid unnatural textures and artifacts in non-RoI regions.
This enables visually plausible reconstruction without degrading machine perception performance.
Additional reconstructed samples are provided in the supplementary material.

\begin{figure}[t]
  \centering
  \includegraphics[width=1\linewidth]{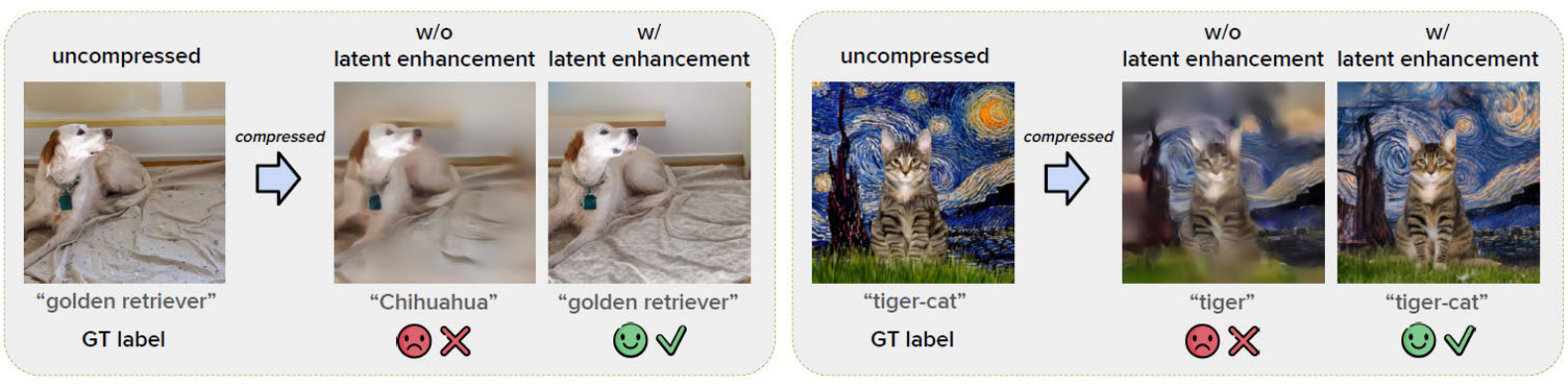}
    \vspace{-2em}

   \caption{Comparison between the SLIM models with and without the latent enhancement.}
  \label{fig: 7_control_to_sample-b}

\end{figure}

\begin{figure}[t]
  \centering
  \includegraphics[width=1\linewidth]{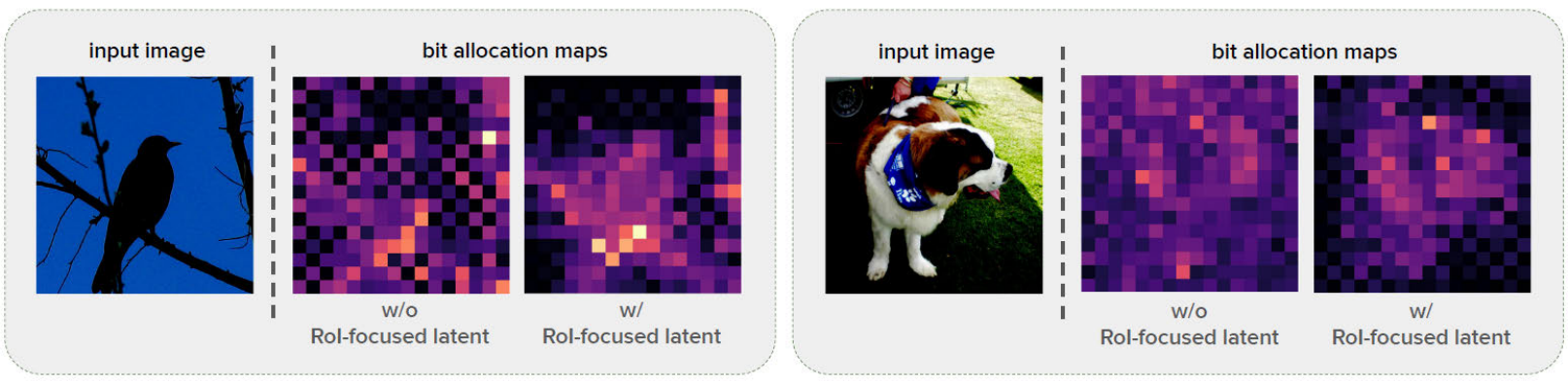}
  \vspace{-2em}
   \caption{Comparison of bit allocation maps with and without the RoI-focused latent objective.}
   \label{fig: 8_bit_allocation_map}

   \vspace{-1em}
\end{figure}

\begin{table}[t]
  \caption{BD-Rate comparison on the ImageNet-val dataset with the ResNet50 classifier.}

  \label{tab: 4_non_roi}
  \centering
  \scalebox{1.0}{
  \begin{tabular}{c c}
    \hline
    \textbf{Model} & \textbf{BD-Rate (\%) $\downarrow$}\\
    \hline

    w/o Latent enhancement & - \\
    w/\,\,\,  Latent enhancement & -67.08 \\
    \hline
    
    w/o RoI-focused latent & - \\
    w/\,\,\,  RoI-focused latent & -18.87 \\
    \hline

    w/o Text-based semantic guidance & - \\
    w/\,\,\,  Text-based semantic guidance & -16.43 \\
    \hline

  \end{tabular}
  }

    \vspace{+0.4em}

\end{table}

\subsection{Ablation Studies}
\label{sec:experiment_ablation}

\noindent \textbf{Latent enhancement.} 
Without the latent enhancement, the decompressed latent obtained from the decompressor is directly decoded by the VAE decoder. 
Then the image reconstructed from such an imperfect latent suffers from degraded perceptual quality and reduced machine vision accuracy as shown in Fig. \ref{fig: 7_control_to_sample-b}.
Additional samples are provided in the supplementary material.
Table \ref{tab: 4_non_roi} further demonstrates that removing latent enhancement substantially degrades the rate-accuracy performance of SLIM.

\noindent \textbf{RoI-focused latent.} 
To confirm that the compressor of SLIM focuses on the RoI areas in the latent space as intended, we observe the bit allocation maps.
Fig. \ref{fig: 8_bit_allocation_map} presents the bit allocation maps of the SLIM models obtained after the training stage 1, comparing models trained with and without the RoI-focused latent objective.
Additional comparison samples are provided in the supplementary material. 
In this comparison, the model without the RoI-focused latent objective uses a modified guide loss computed over the entire latent area instead of the RoI area.
While the bit allocation map of this model exhibits a more uniform bit distribution, the RoI-focused latent objective induces a higher concentration of bits in RoI regions relevant to machine vision. 
As shown in Table \ref{tab: 4_non_roi}, we further confirm that the intended RoI-focused latent improves rate-accuracy performance, demonstrating its crucial role as the source of the RoI-focused spatial cue in semantic latent enhancement.

\noindent \textbf{Text-based semantic guidance.} 
We further verify the effect of text-based semantic guidance, the other source used to construct the dual cues for semantic latent enhancement. 
Specifically, we train a variant of our SLIM model by replacing each caption with an empty string. 
As shown in Table \ref{tab: 4_non_roi}, removing text-based semantic guidance leads to a noticeable performance drop. 
This result demonstrates the importance of this guidance for semantic latent enhancement.

\subsection{Computational Cost}

\begin{table}[t]
    \centering
    \caption{Computational cost of each model for a $512 \times 512$ sized RGB image. All latency values (ms) are obtained under a single NVIDIA RTX 4090 GPU setting.}

    \label{tab: a1_cost}
    \scalebox{1.0}{
    \begin{tabular}{lcc}
        \hline
        Model & Encoding (GFLOPs / ms) & Decoding (GFLOPs / ms) \\
        \hline
        SA-ICM \cite{saicm} & 204.60 / 52.89 & 262.51 / 25.13  \\
        MPA \cite{mpa} & 45.68 / 20.88 & 80.69 / 84.69 \\
        TransTIC \cite{transtic} & 83.94 / 54.96 & 49.11 / 51.71 \\
        AdaptICMH \cite{adapticmh} & 38.64 / 36.20 & 50.11 / 39.77 \\
        OSCAR \cite{oscar} & 564.10 / 47.88 & 1591.33 / 182.69 \\
        SLIM (Ours) & 575.61 / 50.60 & 1082.22 / 212.95 \\
        \hline
    \end{tabular}
    }
\end{table}

We analyze the computational cost of our SLIM model. 
As shown in Table \ref{tab: a1_cost}, the two diffusion-based methods, OSCAR \cite{oscar} and our method, exhibit noticeably longer decoding time than the other baselines. 
The iterative diffusion process in the decoder is likely a major source of the increased computational cost. 
Together with the results in Fig. \ref{fig: 5_bpp_vs_acc} and Tables \ref{tab: 1_bpp_vs_acc}-\ref{tab: 1_extra_bpp_vs_coco}, this observation indicates that our method achieves improved rate-task performance at the cost of increased computational complexity compared with non-diffusion-based approaches.
In addition, compared to the diffusion-based baseline OSCAR, our method requires only about two-thirds of its decoding GFLOPs, indicating lower decoding complexity. 
The overall complexity could be further reduced by leveraging recent techniques for accelerating diffusion model inference (e.g., token merging \cite{tome}).

% ---------------------------------------------------------------
% ---------------------------------------------------------------

%%%%%%%%%%%%%%%%%%%%%%%%%%%%%%%%%%%%%%%%%%%%%%%%%%%%%%%%%%%%%%%%%

% ---------------------------------------------------------------
% ---------------------------------------------------------------

\section{Conclusion}
\label{sec:conclusion}

In this paper, we proposed a novel semantics-based framework of LIC for machine vision, named SLIM.
Our framework is trained with a latent compressor that recovers RoI-focused latents at inference time without requiring explicit masks. These latents are encouraged to emphasize regions that are semantically important for downstream machine tasks.
To further preserve scene-level semantics, we introduced semantic latent enhancement, which leverages both the RoI-focused latent representation and text-based semantic guidance as the sources of two complementary cues.
By effectively exploiting a pretrained diffusion prior, our framework refines decompressed latents prior to image decoding, enabling the reconstruction of high-quality images that are better aligned with machine perception.
Experimental results showed that our SLIM model outperforms the other existing LIC models across diverse downstream tasks. 
In future work, we aim to further reduce the computational cost of our model by leveraging recent training-free acceleration techniques.

% ---------------------------------------------------------------
% ---------------------------------------------------------------

%%%%%%%%%%%%%%%%%%%%%%%%%%%%%%%%%%%%%%%%%%%%%%%%%%%%%%%%%%%%%%%%%

% ---------------------------------------------------------------
% ---------------------------------------------------------------

\section{Acknowledgments}
\label{sec:acknowledgments}

Hyeonjin Lee and Jong-Seok Lee were supported by the National Research Foundation of Korea (NRF) grant (No. RS-2024-00453301, RS-2025-00517159) and by Institute of Information communications Technology Planning Evaluation (IITP) grant (IITP-2025-RS-2024-00428780).
Jun-Hyuk Kim was supported by the Institute of Information \& Communications Technology Planning \& Evaluation (IITP) grant funded by the Korea government (MSIT) [RS-2021-II211341, Artificial Intelligence Graduate School Program (Chung-Ang University)]

% ---- Bibliography ----

\bibliographystyle{splncs04}
\bibliography{main}

%%%%%%%%%%%%%%%%%%%%%%%%%%%%%%%%%%%%%%%%%%%%%%%%%%%%%%%%%%%%%%%%%%%%%%%

%%%%%%%%%%%%%%%%%%%%%%%%%%%%%%%%%%%%%%%%%%%%%%%%%%%%%%%%%%%%%%%%%%%%%%%

%%%%%%%%%%%%%%%%%%%%%%%%%%%%%%%%%%%%%%%%%%%%%%%%%%%%%%%%%%%%%%%%%%%%%%%%%%%%%%

\newpage

% Supplementary numbering
\setcounter{section}{0}
\setcounter{figure}{0}
\setcounter{table}{0}
\setcounter{equation}{0}

\renewcommand{\thesection}{S\arabic{section}}
\renewcommand{\thefigure}{S\arabic{figure}}
\renewcommand{\thetable}{S\arabic{table}}
\renewcommand{\theequation}{S\arabic{equation}}

\newcounter{suppbibitem}
\let\origbibitem\bibitem

\renewcommand{\bibitem}[2][]{
  \stepcounter{suppbibitem}
  \origbibitem[S\arabic{suppbibitem}]{#2}
}

\pagestyle{empty}

\title{SLIM: Semantic-based Low-bitrate Image Compression for Machines\\[0.5em]{\large Supplementary Material}}

\titlerunning{Supplementary Material}

\author{}
\institute{}

\maketitle

\section{Example samples of training dataset}

\begin{figure}[h]
  \centering
  \includegraphics[width=0.65\linewidth]{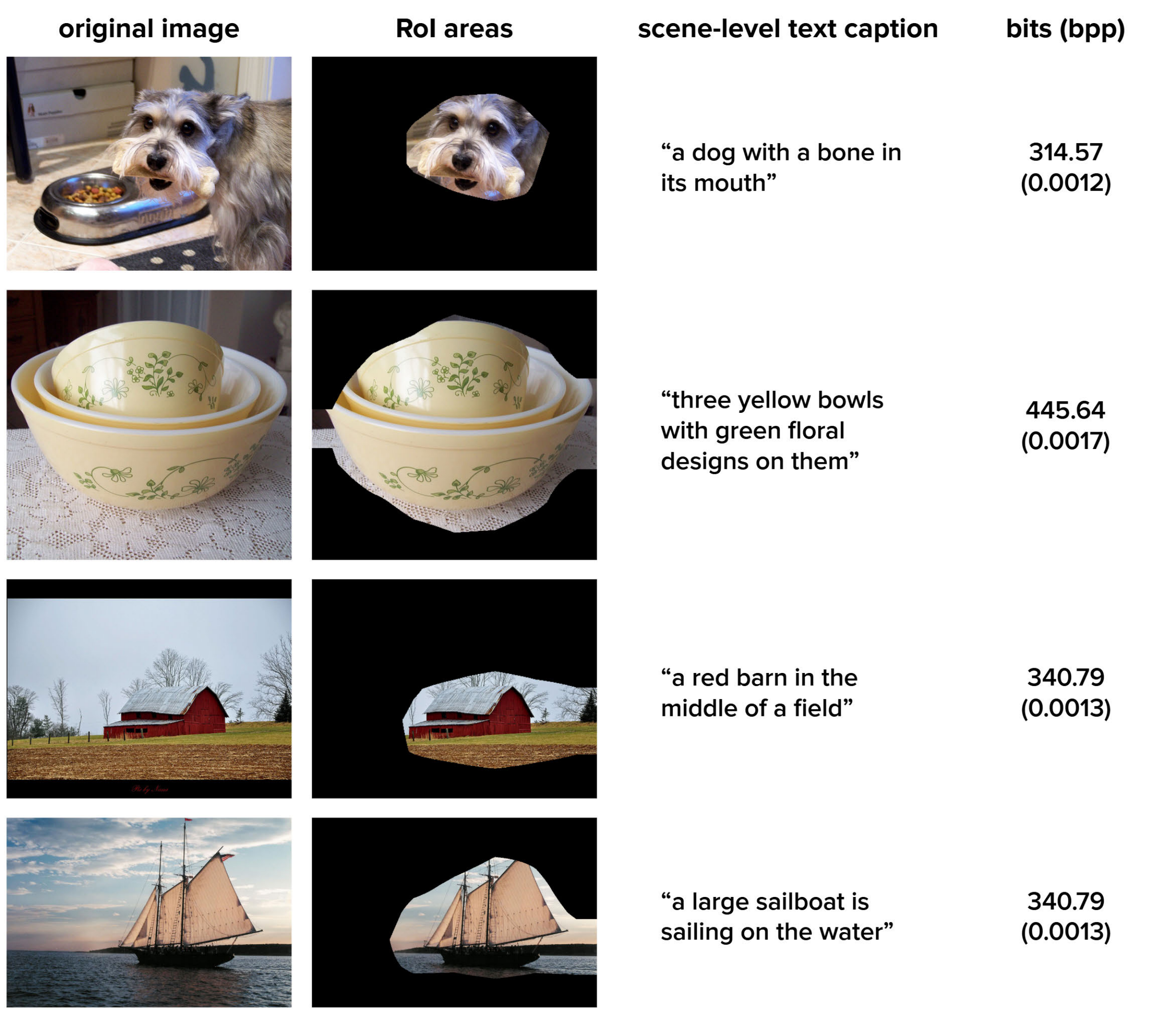}
  \caption{Samples of the RoI areas and text captions for the ImageNet-val dataset.}
  \label{fig: a5_text_caption}
\end{figure}

Fig. \ref{fig: a5_text_caption} shows sample images from the ImageNet-val \cite{imagenet} dataset, along with their corresponding RoI visualizations, scene-level text captions, and compressed caption sizes in bits and bpp.
The RoI-masks are obtained by first computing Grad-CAM \cite{gradcam} heatmaps using a pretrained ResNet-50 classifier \cite{resnet}, normalizing the heatmaps to the range $[0,1]$, and then binarizing them with a threshold of $0.5$.

The scene-level captions are generated by feeding each image into BLIP2 \cite{blip2}, which produces a description of the entire image.
After lossless compression with zlib, the text captions require only 367 bits on average. 
For comparison with the bitrate of compressed images, we normalize the caption bits by the number of pixels in the image ($512\times512=262,144$ pixels). 
On this scale, the captions require only 0.0014 bpp on average, which is extremely small compared with the bitrate required for the compressed image data.

\section{Details of compressor-decompressor pair}
\begin{figure}[h]
  \centering
  \includegraphics[width=1\linewidth]{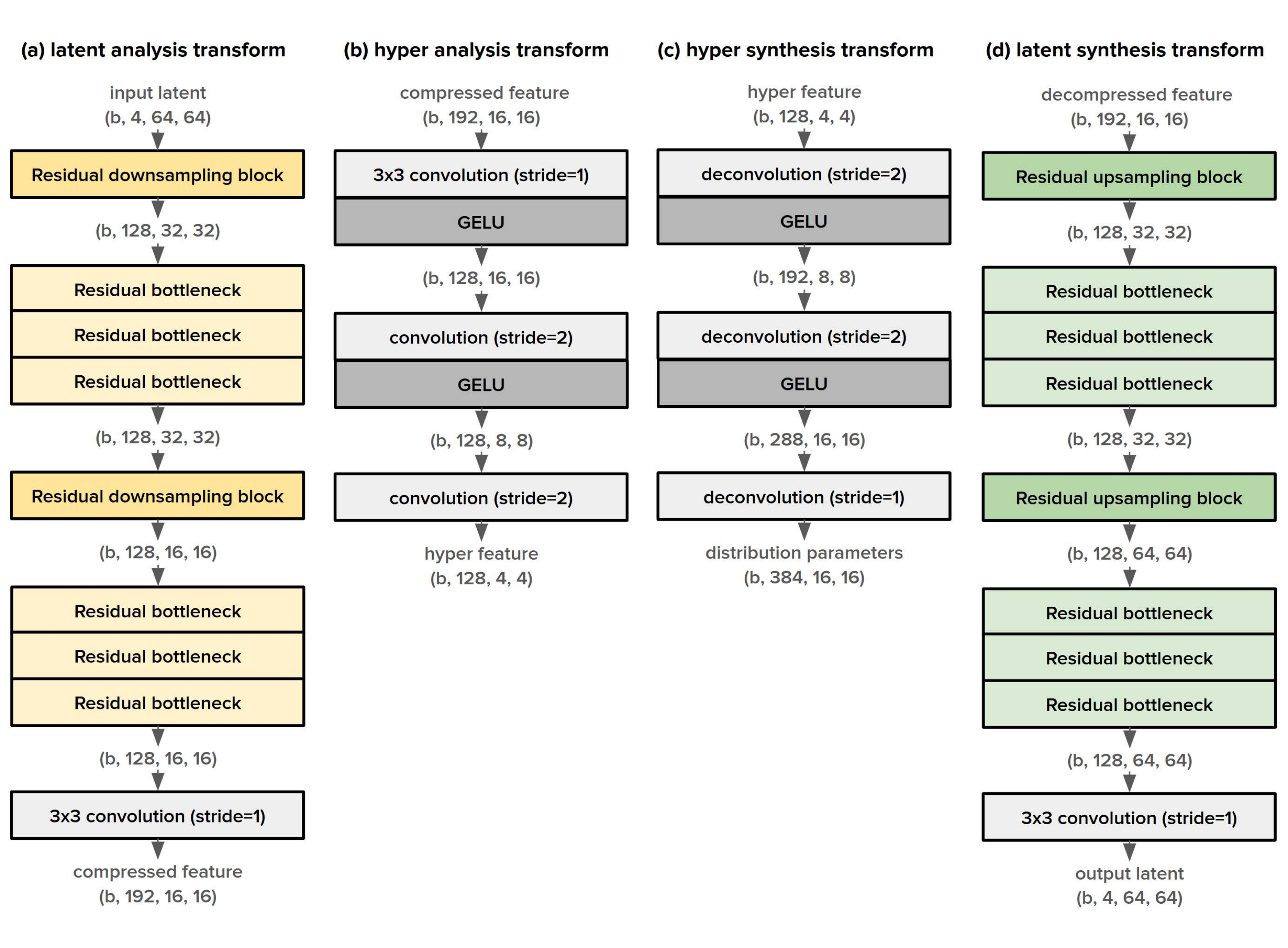}
  \caption{Architecture of the compressor-decompressor pair of SLIM. Each residual bottleneck consists of three convolution layers and two GELU layers with a residual connection, whereas the residual downsampling and upsampling blocks consists of three convolution layers and two leaky ReLU layers, also with a residual connection.}
  \label{fig: a1_details}
\end{figure}

Inspired by the ELIC \cite{elic} framework for image compression, the compressor-decompressor pair of our SLIM applies hyperprior-based compression on the latent produced by the VAE encoder, using a checkerboard slicing strategy. The pair consists of a latent analysis transform, hyper analysis transform, hyper synthesis transform, and latent synthesis transform. 
The detailed structure of the pair models is shown in Fig. \ref{fig: a1_details}.

The compressor first applies the latent analysis transform to the input latent, producing the compressed features.
Then, the hyper analysis transform extracts hyper features from these compressed features to support entropy coding.
During entropy-coding, the compressed features are divided into multiple slices, which are processed sequentially using a checkerboard-channel context.

On the decompressor side, the hyper synthesis transform produces the distribution parameters from the hyper features, leveraging the checkerboard-channel context to better estimate those parameters for entropy-decoding each slice of the compressed features.
Finally, the latent synthesis transform reconstructs the output latent from the decompressed features.

%%%%%%%%%%%%%%%%%%%%%%%%%%%%%%%%%%%%%%%%%%%%%%%%%%%%%%%%%%%%%%%%%

\section{Caption bitrate}
\begin{figure}[H]
  \centering

  \begin{subfigure}{0.48\linewidth}
    \centering
    \includegraphics[width=\linewidth]{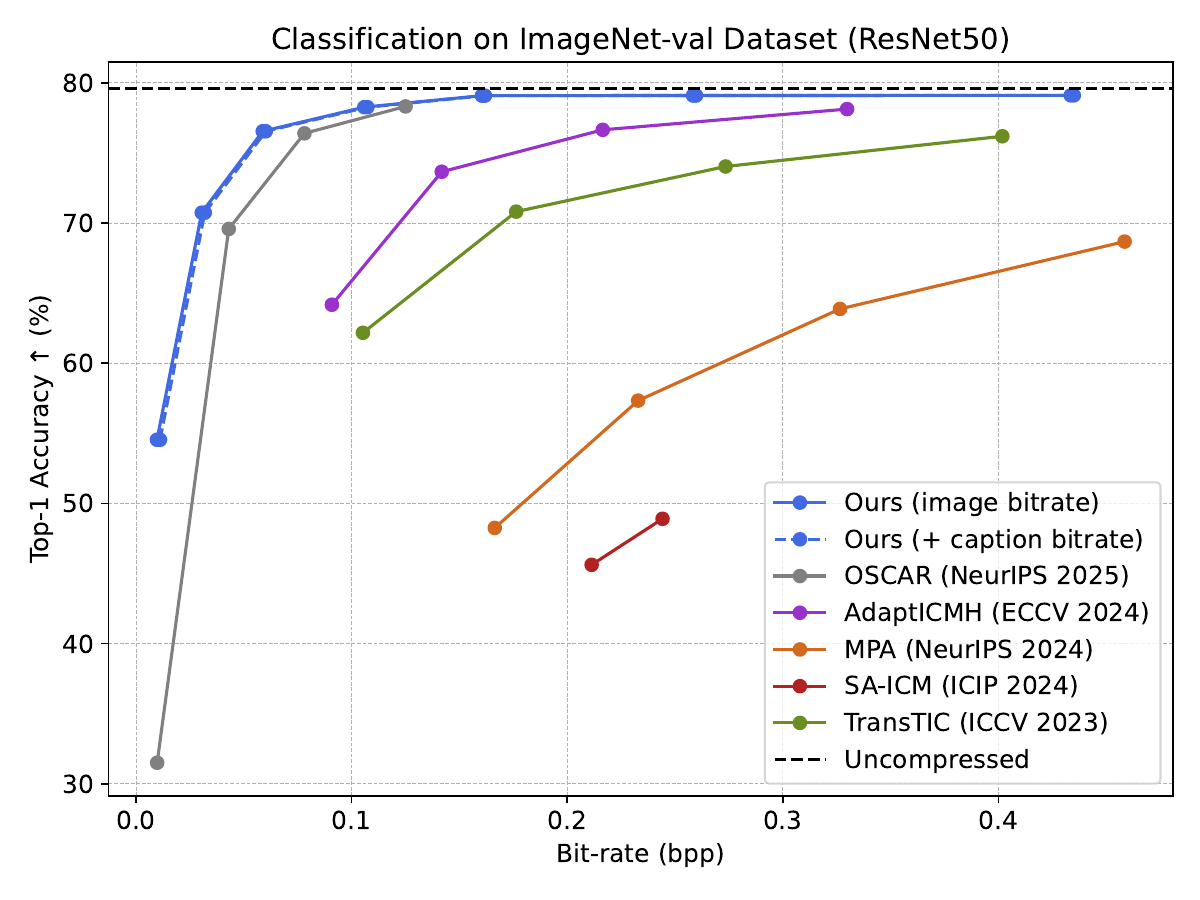}
  \end{subfigure}
  \begin{subfigure}{0.48\linewidth}
    \centering
    \includegraphics[width=\linewidth]{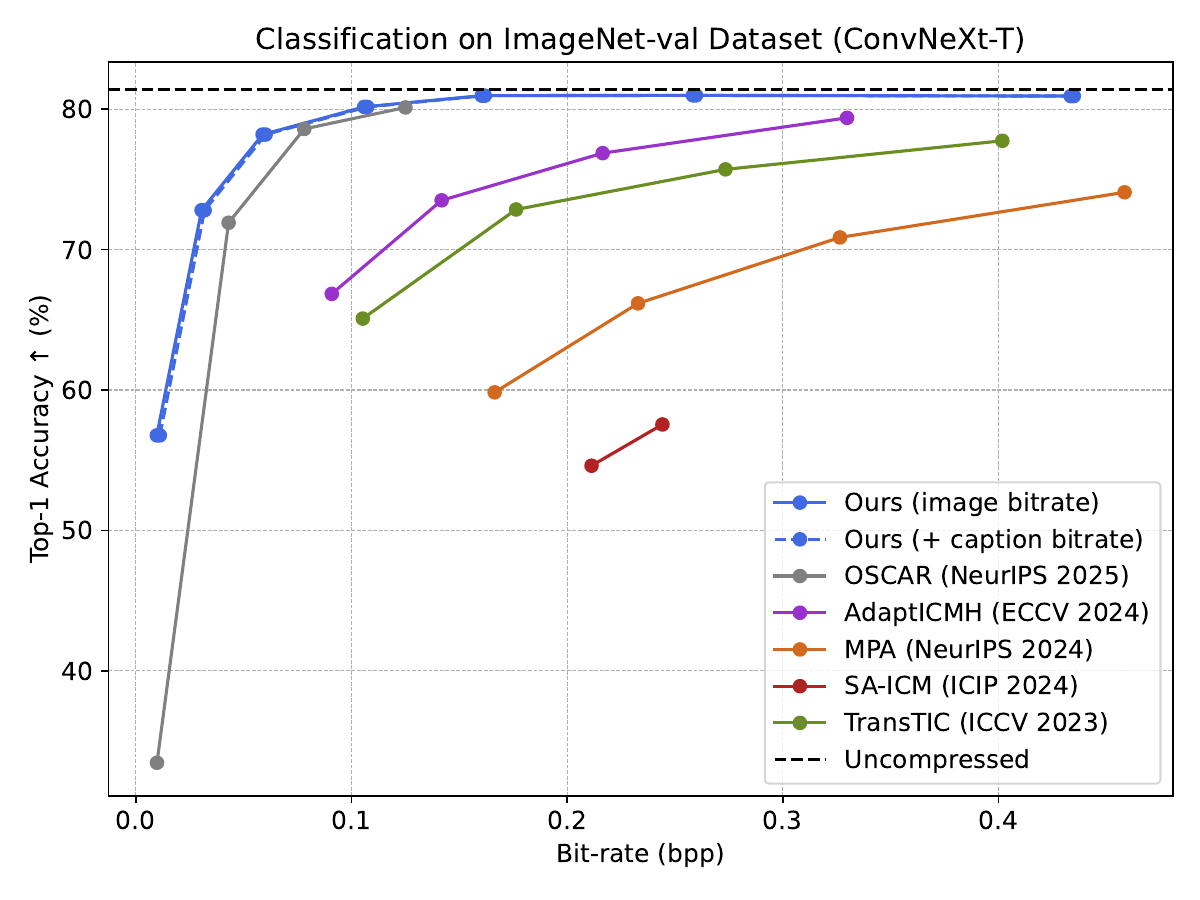}
  \end{subfigure}

  \vspace{0.5em}

  \begin{subfigure}{0.48\linewidth}
    \centering
    \includegraphics[width=\linewidth]{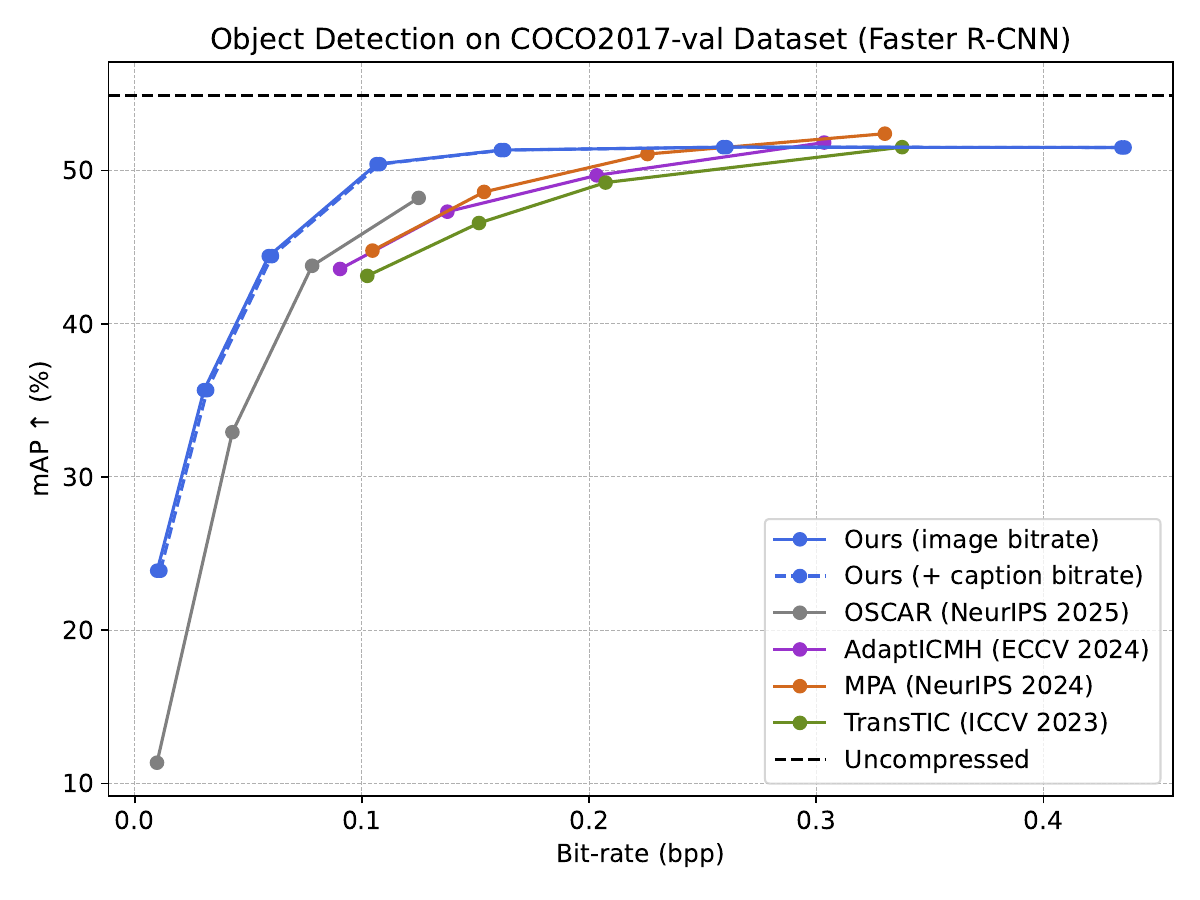}
  \end{subfigure}
  \begin{subfigure}{0.48\linewidth}
    \centering
    \includegraphics[width=\linewidth]{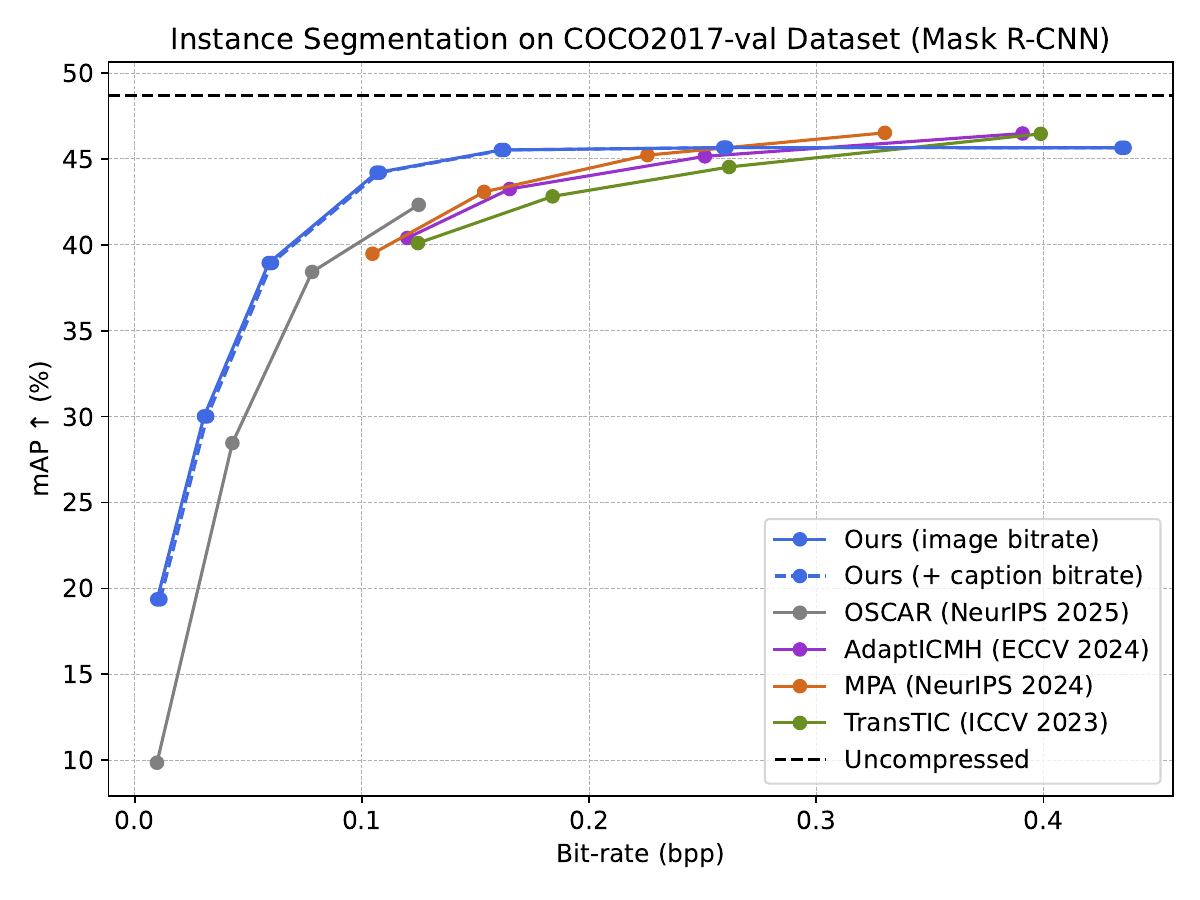}
  \end{subfigure}

  \caption{
    Rate-task performance with caption overhead included.
    Dashed curves denote our method including caption bitrate.
  }
  \label{fig:bpp1}
\end{figure}

This section presents the results including the caption bitrate. Fig. \ref{fig:bpp1} and Table \ref{tab:bpp1} show the rate-task performance and BD-Rate results, respectively, with an additional $0.0014$ bpp included for caption transmission. Our SLIM model still outperforms the baselines after accounting for this additional bitrate.

\begin{table}[H]
  \centering
  \caption{
    BD-Rate (\%) comparison with caption overhead included.
  }
  \label{tab:bpp1}

  \footnotesize
  \setlength{\tabcolsep}{4pt}
  \resizebox{\linewidth}{!}{
  \begin{tabular}{l c c c c}
    \hline
    \textbf{Method}
    & \textbf{ResNet50 $\downarrow$}
    & \textbf{ConvNeXt-T $\downarrow$}
    & \textbf{Faster R-CNN $\downarrow$}
    & \textbf{Mask R-CNN $\downarrow$} \\
    \hline
    MPA \cite{mpa} & -- & -- & -- & -- \\
    TransTIC \cite{transtic} & -67.23 & -55.16 & 25.40 & 25.22 \\
    AdaptICMH \cite{adapticmh} & -75.60 & -65.14 & 7.75 & 8.17 \\
    OSCAR \cite{oscar} & -90.89 & -87.26 & -16.15 & -12.94 \\
    \textbf{Ours}
    & \textbf{-94.95}
    & \textbf{-92.68}
    & \textbf{-43.65}
    & \textbf{-40.53} \\
    \textbf{Ours (+ caption bitrate)}
    & \textbf{-94.46}
    & \textbf{-92.11}
    & \textbf{-42.68}
    & \textbf{-39.55} \\
    \hline
  \end{tabular}
  }
\end{table}

%%%%%%%%%%%%%%%%%%%%%%%%%%%%%%%%%%%%%%%%%%%%%%%%%%%%%%%%%

\section{Additional Object Detection Metrics}

\begin{figure}[H]
  \centering

  \begin{subfigure}{0.48\linewidth}
    \centering
    \includegraphics[width=\linewidth]{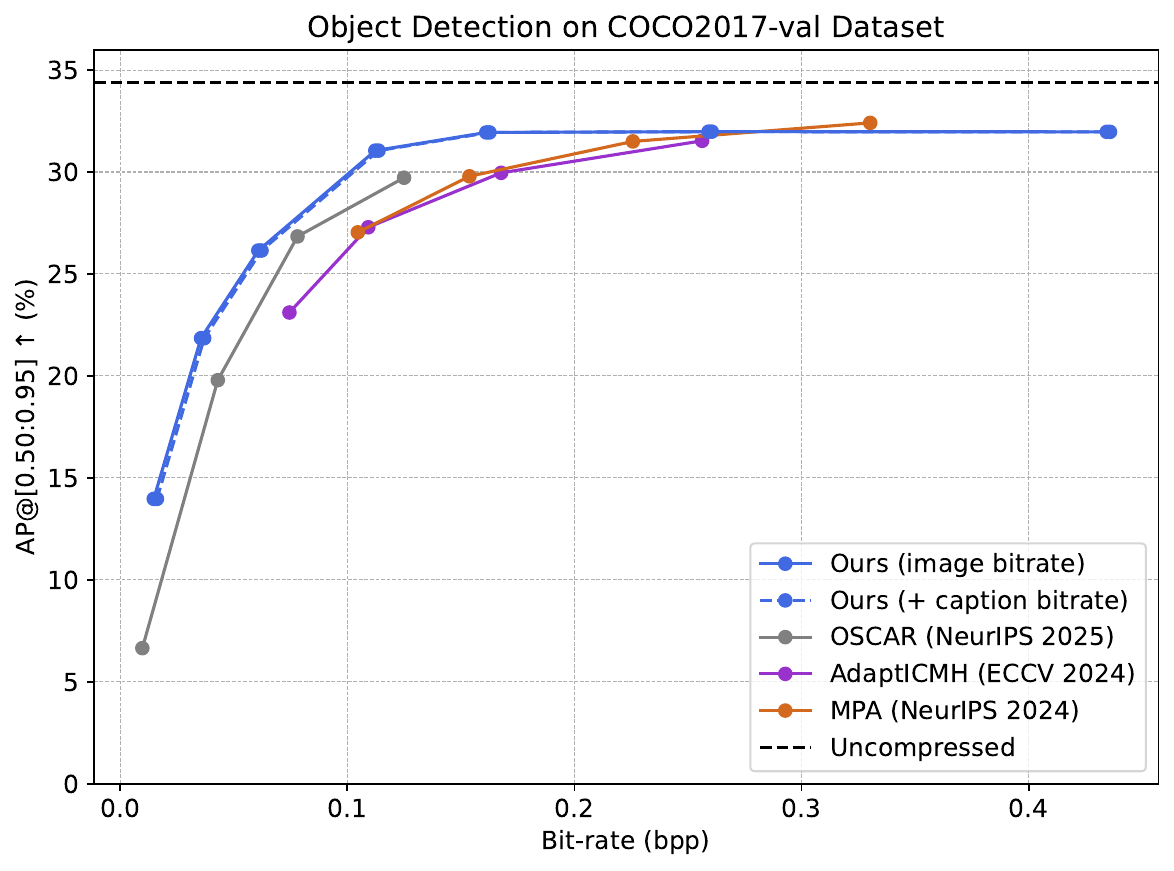}
  \end{subfigure}
  \begin{subfigure}{0.48\linewidth}
    \centering
    \includegraphics[width=\linewidth]{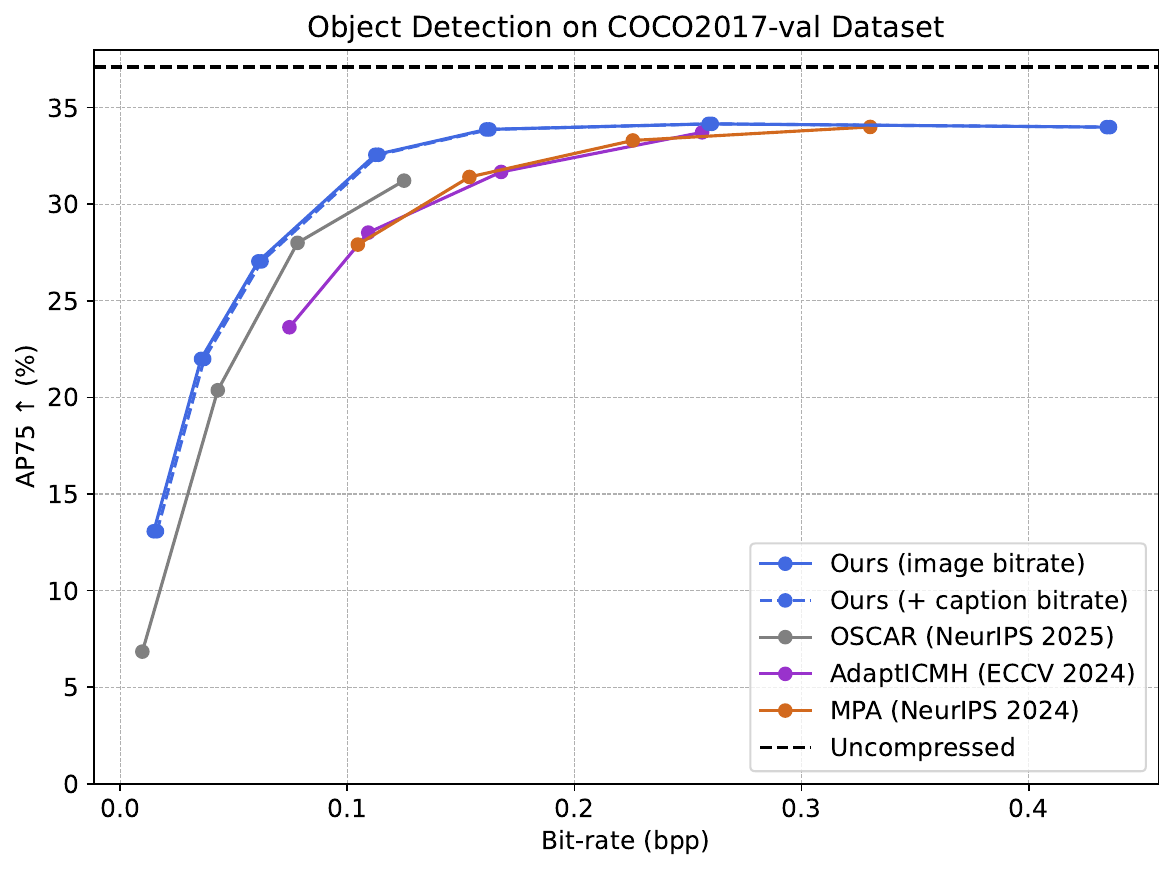}
  \end{subfigure}

  \caption{
    Rate-performance comparison on the COCO2017-val dataset using Faster R-CNN for object detection, in terms of AP@[0.50:0.95] and AP75.
  }
  \label{fig:coco_ap}
\end{figure}

This section provides additional evaluation results for object detection on the COCO2017-val \cite{coco} dataset using Faster R-CNN \cite{fasterrcnn}, following the same experimental setting as in the main paper. The mAP metric at IoU = 0.5 reported in the main paper corresponds to AP50. Fig. \ref{fig:coco_ap} additionally shows the rate-performance results using AP@[0.50:0.95] and AP75.

%%%%%%%%%%%%%%%%%%%%%%%%%%%%%%%%%%%%%%%%%%%%%%%%%%%%%%%%%

\section{Additional task evaluation}

This section provides additional evaluation results for human keypoint detection using images from the COCO2017-val dataset. We evaluate the reconstructed images using Keypoint R-CNN, based on the keypoint detection framework of Mask R-CNN \cite{maskrcnn}. Table \ref{table:general} shows the corresponding BD-Rate results. Our SLIM model consistently outperforms the baselines on this additional task.

\begin{table}[H]
  \centering
  \caption{
    BD-Rate comparison on the COCO2017-val dataset using Keypoint R-CNN for human keypoint detection.
  }
  \label{table:general}

  \begin{tabular}{l c}
    \hline
    \textbf{Method}
    & \textbf{BD-Rate (\%) $\downarrow$} \\
    \hline
    MPA \cite{mpa} & -- \\
    AdaptICMH \cite{adapticmh} & -8.11 \\
    OSCAR \cite{oscar} & -8.77 \\
    \textbf{Ours} & \textbf{-16.29} \\
    \textbf{Ours (+ caption bitrate)} & \textbf{-15.14} \\
    \hline
  \end{tabular}
\end{table}

%%%%%%%%%%%%%%%%%%%%%%%%%%%%%%%%%%%%%%%%%%%%%%%%%%%%%%%%%

\newpage

\section{Additional reconstructed images}
This section presents further qualitative results of our model. Fig. \ref{fig: a2_recon_samples} shows reconstructed images from the ImageNet-val dataset at comparable bpp levels.

\begingroup
\setlength{\intextsep}{10pt}

\begin{figure}[H]
  \centering
  \includegraphics[width=1.0\linewidth]{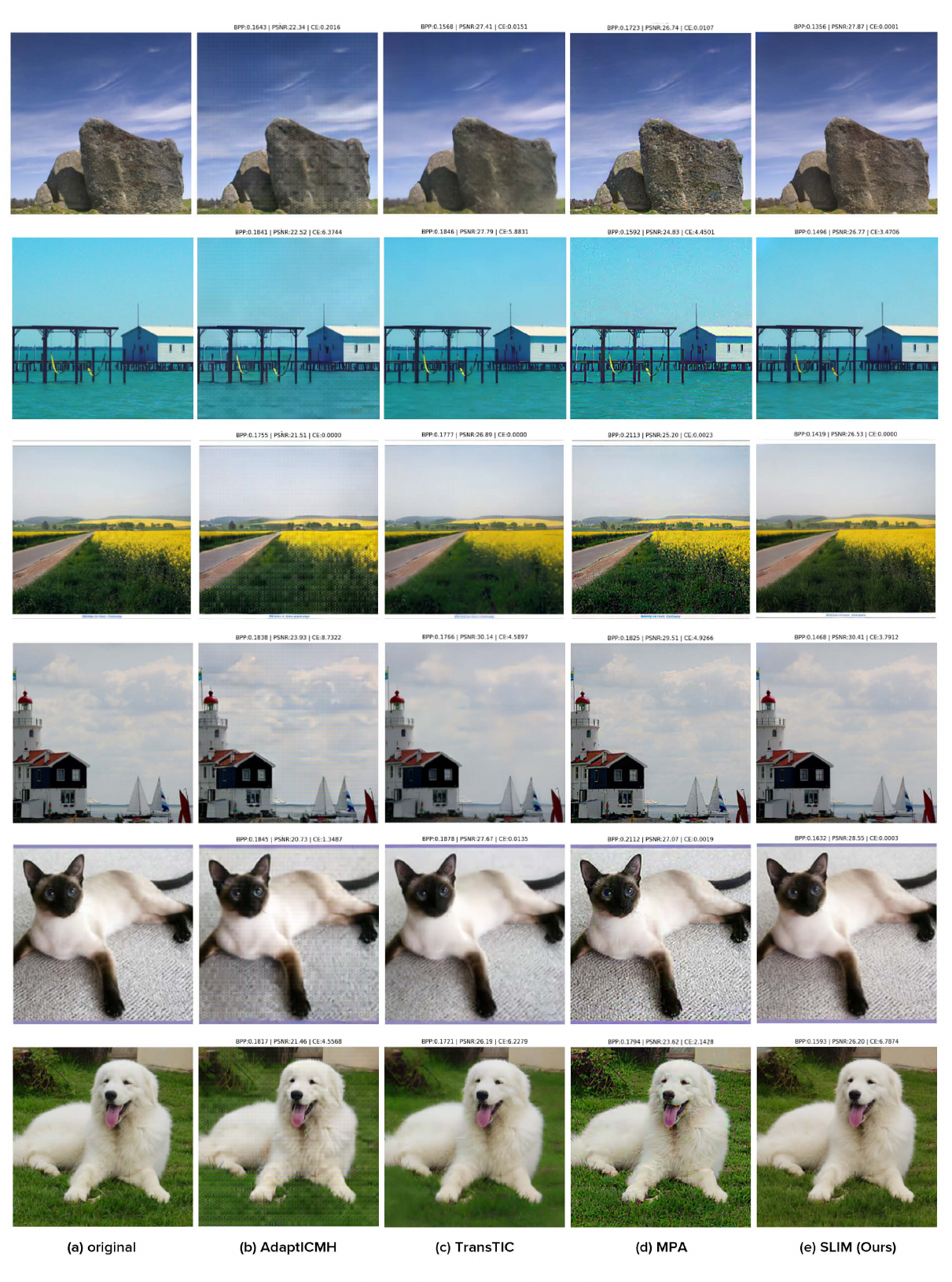}
  \caption{Reconstructed images from our SLIM model at comparable bpp levels.}
  \label{fig: a2_recon_samples}
\end{figure}

\endgroup

\section{Additional samples from ablation studies}
\begin{figure}[H]
  \centering
  \begin{subfigure}{0.49\linewidth}
  \centering
    \includegraphics[width=0.91\linewidth]{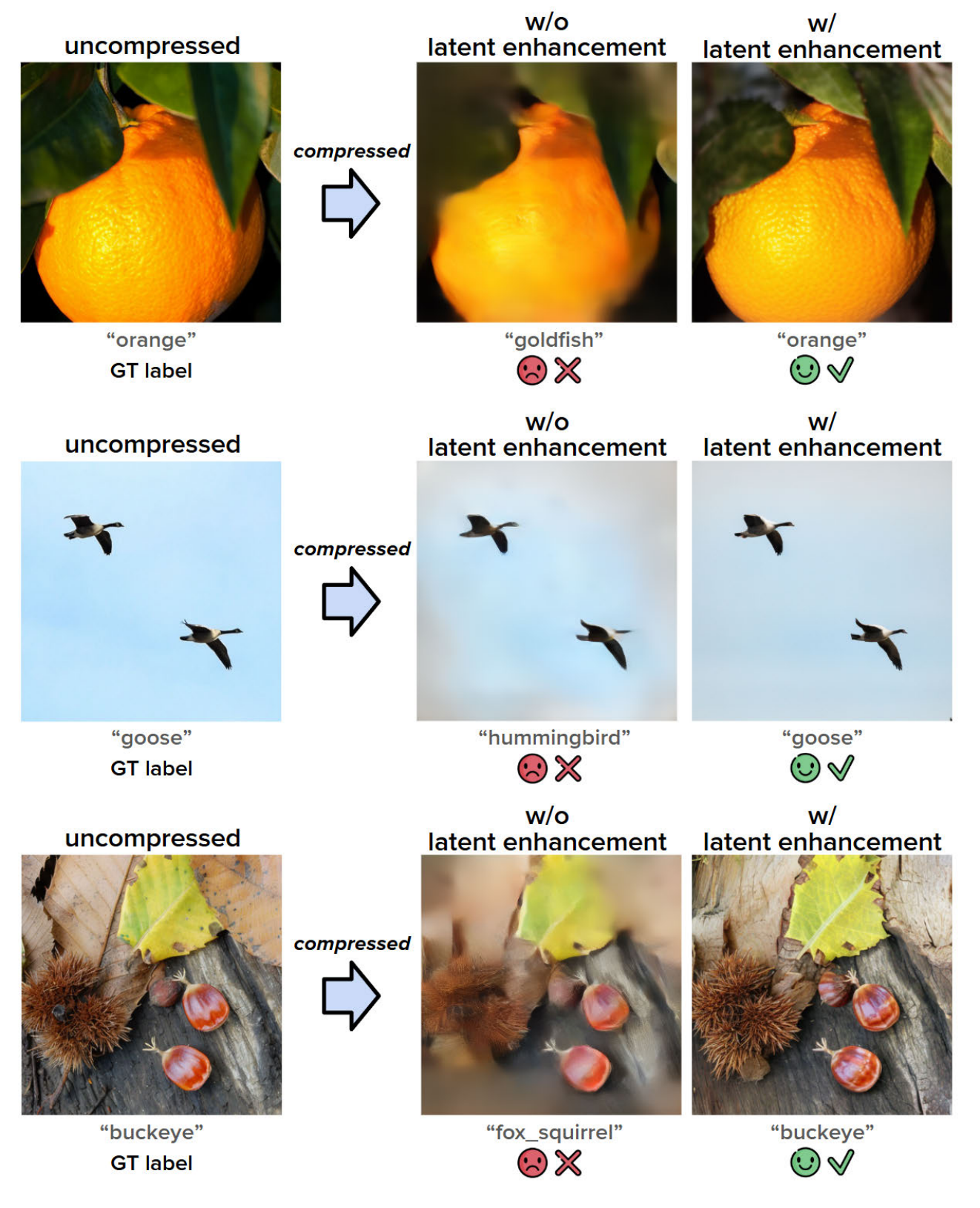}
    \caption{Reconstructed images with and without the latent enhancement}
    \label{fig: a3_1_without_diffusion}
  \end{subfigure}
  \hfill
  \begin{subfigure}{0.49\linewidth}
  \centering
    \includegraphics[width=0.88\linewidth]{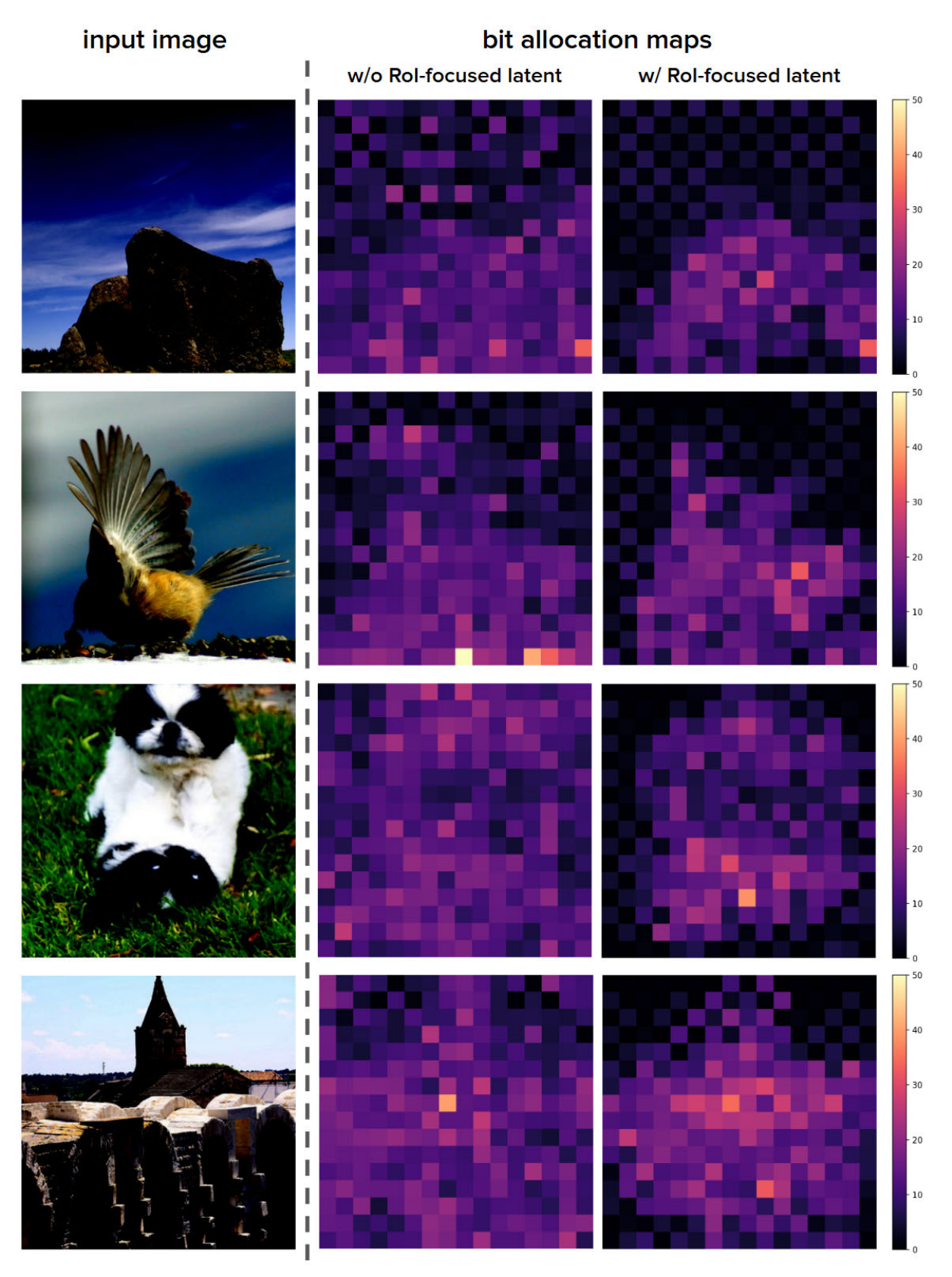}
    \caption{Bit allocation maps with and without the RoI-focused latent objective}
    \label{fig: a3_2_bit_allocation_map}
  \end{subfigure}
  \caption{Comparison between the SLIM models with and without (a) latent enhancement, and (b) RoI-focused latent objective.}
\end{figure}

This section provides additional samples for the ablation studies.
Fig. \ref{fig: a3_1_without_diffusion} shows reconstructed images comparing the SLIM models with and without the latent enhancement, and Fig. \ref{fig: a3_2_bit_allocation_map} shows the bit allocation maps obtained from the SLIM models with and without the RoI-focused latent objective.

%%%%%%%%%%%%%%%%%%%%%%%%%%%%%%%%%%%%%%%%%%%%%%%%%%%%%%%%%

\end{document}